\documentclass[a4paper,fleqn,usenatbib]{mnras}

\usepackage{newtxtext,newtxmath}

\usepackage[T1]{fontenc}
\usepackage{ae,aecompl}

\usepackage{graphicx}	
\usepackage{amsmath}	
\usepackage{amssymb}	

\newcommand{\elecd}{$N_{\rm e}$}
\newcommand{\elect}{$T_{\rm e}$}
\newcommand{\tf}{$t^{2}$}
\newcommand{\hb}{H$\beta$}
\newcommand{\ha}{H$\alpha$}

\newcommand{\foiii}{[O\thinspace{\sc iii}]}

\newcommand{\foi}{[O\thinspace{\sc i}]}
\newcommand{\foii}{[O\thinspace{\sc ii}]}
\newcommand{\fsii}{[S\thinspace{\sc ii}]}
\newcommand{\fsiii}{[S\thinspace{\sc iii}]}
\newcommand{\fni}{[N\thinspace{\sc i}]}
\newcommand{\fnii}{[N\thinspace{\sc ii}]}
\newcommand{\fariv}{[Ar\thinspace{\sc iv}]}
\newcommand{\fcliii}{[Cl\thinspace{\sc iii}]}

\newcommand{\fneiii}{[Ne\thinspace{\sc iii}]}

\newcommand{\ffeii}{[Fe\thinspace{\sc ii}]}
\newcommand{\ffeiii}{[Fe\thinspace{\sc iii}]}

\newcommand{\nitroi}{N\thinspace{\sc i}}
\newcommand{\nii}{N\thinspace{\sc ii}}

\newcommand{\silii}{Si\thinspace{\sc ii}}
\newcommand{\oi}{O\thinspace{\sc i}}
\newcommand{\oii}{O\thinspace{\sc ii}}

\newcommand{\cii}{C\thinspace{\sc ii}}

\newcommand{\fariii}{[Ar\thinspace{\sc iii}]}

\newcommand{\hi}{H\,{\sc i}}
\newcommand{\hii}{H\thinspace{\sc ii}}

\newcommand{\hei}{He\thinspace{\sc i}}

\newcommand\ionic[2]{${\rm #1^{#2}}$}           

\newcommand{\cmc}{{\rm cm$^{-3}$}}

\title[Oxygen gradient at the Galactic anticentre]{The radial abundance gradient of oxygen towards the Galactic anticentre}

\author[C. Esteban et al.]{
C. Esteban,$^{1, 2}$\thanks{E-mail: cel@iac.es}
X. Fang,$^{3, 4}$
J. Garc{\'{\i}}a-Rojas$^{1, 2}$
and L. Toribio San Cipriano$^{1, 2}$
\\
$^{1}$Instituto de Astrof\'isica de Canarias, E-38200 La Laguna, Tenerife, Spain\\
$^{2}$Departamento de Astrof\'isica, Universidad de La Laguna, E-38206, La Laguna, Tenerife, Spain\\
$^{3}$Laboratory for Space Research, Faculty of Science, University of Hong Kong, Pokfulam Road, Hong Kong, China\\
$^{4}$Department of Earth Sciences, Faculty of Science, University of Hong Kong, Pokfulam Road, Hong Kong, China
}

\date{Accepted XXX. Received YYY; in original form ZZZ}

\pubyear{2015}

\begin{document}
\label{firstpage}
\pagerange{\pageref{firstpage}--\pageref{lastpage}}
\maketitle

\begin{abstract}
We present deep optical spectroscopy of eight {\hii} regions located 
in the anticentre of the Milky Way.  The spectra were obtained at the 
10.4m GTC and 8.2m VLT.  We determined \elect({\fnii}) for all objects and 
\elect({\foiii}) for six of them. 
We also included in our analysis an additional sample of 13 inner-disc 
Galactic {\hii} regions from the literature that have excellent {\elect} 
determinations. 
We adopted the same methodology and atomic dataset to determine the 
physical conditions and ionic abundances for both samples. 
We also detected the {\cii} and {\oii} optical recombination lines in 
Sh~2-100, which enables determination of the abundance discrepancy factor 
for this object. We found that the slopes of the radial oxygen gradients 
defined by the {\hii} regions from $R_{25}$ (= 11.5~kpc) to 17~kpc and 
those within $R_{25}$ are similar within the uncertainties, indicating the 
absence of flattening in the radial oxygen gradient in the outer Milky Way. 
In general, we found that the scatter of the O/H ratios of {\hii} regions 
is not substantially larger than the observational uncertainties.  The largest 
possible local inhomogeneities of the oxygen abundances are of the order of 
0.1~dex.  We also found positive radial gradients in \elect({\foiii}) 
and \elect({\fnii}) across the Galactic disc. The shapes of these temperature gradients 
are similar and also consistent with the absence of flattening of the metallicity 
distribution in the outer Galactic disc. 
\end{abstract}




\section{Introduction}
\label{sec:intro}

The spatial distributions of abundances of chemical elements along galactic discs -- the radial gradients -- have been derived for many galaxies and are key to understanding the chemical evolution of galaxies. These gradients reflect the star formation history as well as the effects of gas flows in stellar systems. {\hii} regions are fundamental probes to trace the present-day composition of the gas-phase interstellar medium. They have been widely used to determine the radial abundance gradients -- especially for O, a proxy of metallicity in the analysis of ionised gaseous nebula -- in the Milky Way and other spiral galaxies. Some authors have claimed -- based on both the {\hii} region and planetary nebula data -- that radial abundance gradients may flatten at the outer parts of the Milky Way disc \citep[e. g.][]{fichsilkey91, vilchezesteban96, macieletal06}, although other authors do not support such claims \citep[e.g.][]{caplanetal00, henryetal10}. Metallicity determinations of other kinds of objects, such as cepheids \citep[e.g.][]{lucketal03, pedicellietal09, lemasleetal13} and Galactic open clusters \citep[e.g.][]{yongetal12}, find in general some hints of such flattening but some of them do not. 

\begin{table*}
   \centering
   \caption{Journal of observations.}
   \label{tab:journal}
   \begin{tabular}{lcccccccc}
        \hline
        & R.A.$^{\rm a}$ & Decl.$^{\rm a}$ & $R_\mathrm{G}^{\rm b}$ & Telescope/& PA & Extracted area & Grating or & Exposure time \\
        {\hii} region & (J2000) & (J2000) &  (kpc) &  /Spectrograph & ($^\circ$) & ($\mathrm{arcsec}^2$) &  configuration & (s) \\
        \hline                                                                                                                                                                  
         Sh 2-83    & 19:24:52.55 & $+$20:47:24.4 & 15.3 $\pm$ 0.1 & GTC/OSIRIS & 90 & 38.1 $\times$ 0.8& R1000B & 3 $\times$ 882 \\
         & & & & & & & R2500V & 3 $\times$ 882 \\
         Sh 2-100  & 20:02:00.69 & $+$33:29:23.9 & 9.4 $\pm$ 0.3 & GTC/OSIRIS & $-$55 & 28.5 $\times$ 0.8& R1000B & 3 $\times$ 882 \\
         & & & & & & & R2500V & 3 $\times$ 882 \\  
         Sh 2-127 & 21:28:40.95 & $+$54:42:13.9 & 14.2 $\pm$ 1.0 & GTC/OSIRIS & 1 & 6.9 $\times$ 0.8& R1000B & 3 $\times$ 882 \\
         & & & & & & & R2500V & 3 $\times$ 882 \\
         Sh 2-128 & 21:32:49.86 & $+$55:53:16.2 & 12.5 $\pm$ 0.4 & GTC/OSIRIS & 86 & 24.1 $\times$ 0.8& R1000B & 3 $\times$ 882 \\ 
         & & & & & & & R2500V & 3 $\times$ 882 \\         
         Sh 2-209 & 04:11:25.69 & $+$51:14:33.8 & 17.0 $\pm$ 0.7 & GTC/OSIRIS & 35 & 45.7 $\times$ 0.8& R1000B & 3 $\times$ 882 \\
         & & & & & & & R2500V & 3 $\times$ 882 \\
         Sh 2-212 & 04:40:56.13 & $+$50:26:53.1 & 14.6 $\pm$ 1.4 & GTC/OSIRIS & $-$70 & 38.1 $\times$ 0.8& R1000B & 3 $\times$ 882 \\
         & & & & & & & R2500V & 3 $\times$ 882 \\
         Sh 2-288 & 07:08:48.90 & $+$07:08:48.9 & 14.1 $\pm$ 0.4 & GTC/OSIRIS & $-$32 & 28.5 $\times$ 0.8& R1000B & 3 $\times$ 882 \\
         & & & & & & & R2500V & 3 $\times$ 882 \\
         Sh 2-298 & 07:18:28.10 & $-$13:17:19.7 & 11.9 $\pm$ 0.7 & VLT/UVES & 90 & 8 $\times$ 3 & DIC1(346+580) & 3 $\times$ 540 \\
         & & & & & & & DIC2(437+860) & 3 $\times$ 1620 \\
        \hline
   \end{tabular}
    \begin{description}
      \item[$^{\rm a}$] Coordinates of the slit centre. 
      \item[$^{\rm b}$] Galactocentric distances assuming the Sun at 8 kpc.
    \end{description}
\end{table*}

There is strong evidence for flat radial gradients from the spectroscopic observations of {\hii} regions in the outer discs of some nearby spiral galaxies, as in the cases of M83, NGC~1512 or NGC~3621 \citep{bresolinetal09, bresolinetal12} among others. Moreover, \citet{sanchezetal14} presented a catalogue of {\hii} regions in several hundred galaxies from the CALIFA survey and many of them show an external flattening in oxygen. Flattening or even upturn of the metallicity gradient has also been found from photometry of red giant branch stars in the outer discs of spiral galaxies NGC~300 and NGC~7793 \citep[e. g.][]{vlajicetal11}. \citet{bresolinetal12} and more recently \citet{bresolin16} discuss several mechanisms that can produce such flattening: leveling of the star formation efficiency, stellar migration, radial metal mixing, or enriched gas infall. These authors indicate that flattening of the radial abundance gradients in external spiral galaxies occurs approximately at the isophotal radius, $R_{25}$, which in the case of the Milky Way is about 11.5 kpc \citep{devaucouleurspence78}. Metallicity studies based on Galactic cepheids and open clusters indicate that the change in the slope of the Milky Way occurs at about 9 kpc \citep{lepineetal11}. \citet{estebanetal13} obtained very deep spectra of the bright {\hii} region NGC~2579, which is located at a galactocentric distance of 12.4 kpc, close to $R_{25}$, the best data ever taken for an {\hii} region at the outer Galactic disc. They found that their O/H and C/H ratios are consistent with flattened gradients. Very recently, \cite{fernandezmartinetal16} presented the 4.2m WHT spectroscopic observations of nine {\hii} regions located at $R_{\rm G}$ > 11 kpc, but only detected the temperature-sensitive auroral lines in five of them. In addition, those authors re-analysed the data of other {\hii} regions  retrieved from the literature. The results of \cite{fernandezmartinetal16} do not confirm the presence of flattening at the Galactic anticentre, and they point out the necessity of more observations of {\hii} regions in the outer part of the Galaxy to establish the true shape of the metallicity gradient. 

  \begin{figure*}
   \centering
   \includegraphics[scale=0.22]{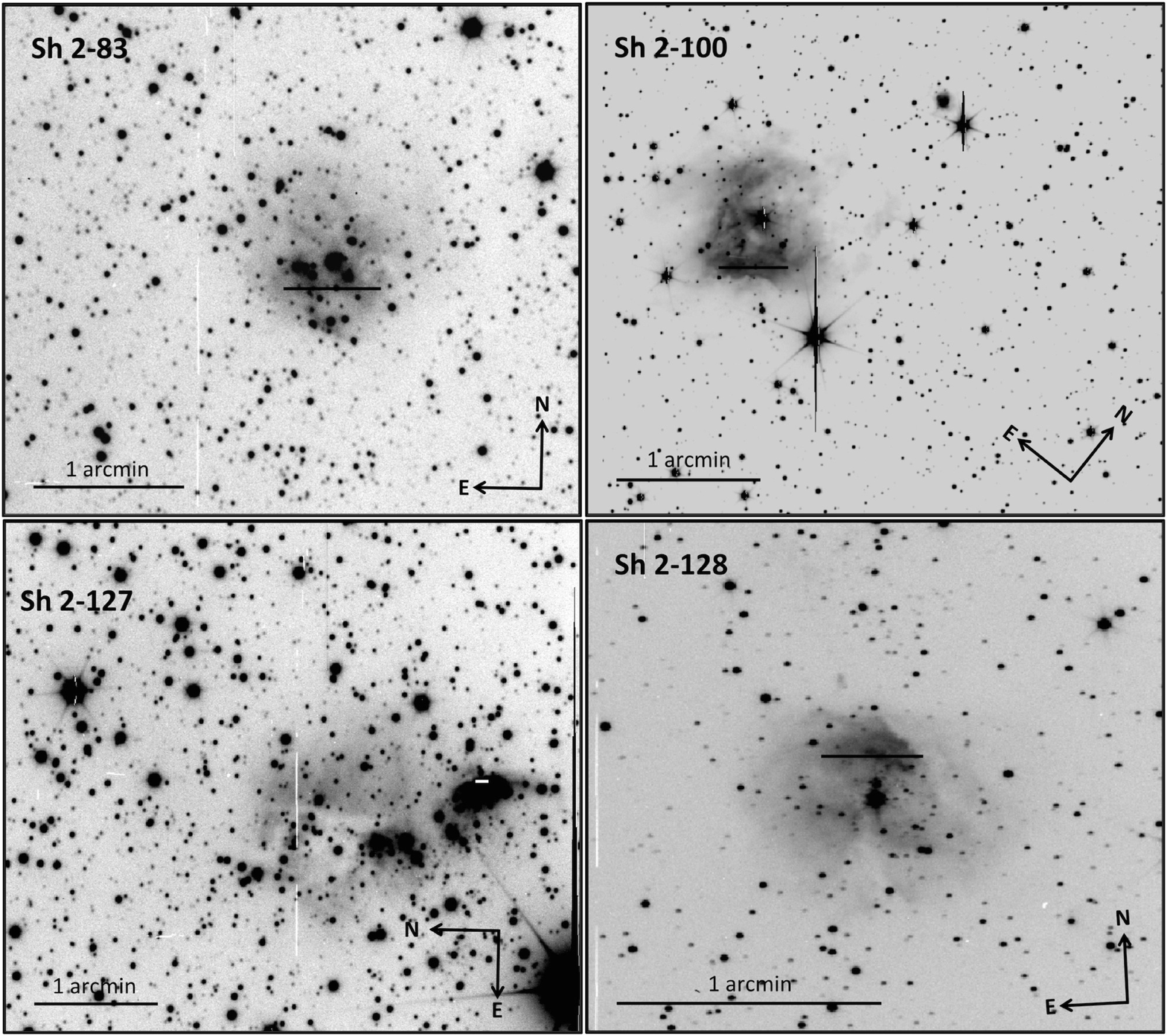} 
   \caption{
   The GTC {\it g}-band acquisition images of Sh~2-83, Sh~2-100, Sh~2-127 and Sh~2-128. The position and length of the aperture extracted for each object are indicated.
   }
   
   \label{fig:slits_1}
  \end{figure*}

The paucity of accurate abundance determinations for {\hii} regions in the anticentre direction has been an enduring problem in the exploration of the shape of the O gradient in the Galactic disc. Those distant nebulae are usually faint and the number of them with direct determinations of electron temperature, {\elect}, is rather limited. A high-quality determination of {\elect} is essential to obtain reliable O abundances. We have carried out a project to obtain very deep spectroscopy of a selected sample of {\hii} regions located close to or beyond $R_{25}$ in order to increase the sample with reliable measurements of the O/H ratios in the direction of Galactic anticentre. We have selected a sample of eight relatively bright objects with $R_{\rm G}$ in the range 9.4--17.0 kpc that show a high ionization degree, $I$({\foiii} $\lambda$5007)/$I$(H$\beta$) $\geq$ 1.0, for ensuring as much as possible the detection of the {\foiii} $\lambda$4363 auroral line. These data are complemented with some other outer-disc {\hii} regions, NGC~2579 \citep{estebanetal13}, Sh~2-311 \citep{garciarojasetal05} and NGC~7635 \citep{estebanetal16}, which were previously observed and published by our group and for which we have excellent determinations of {\elect}.

\section{Observations and Data Reduction}
\label{sec:obs}

  \begin{figure*}
   \centering
   \includegraphics[scale=0.22]{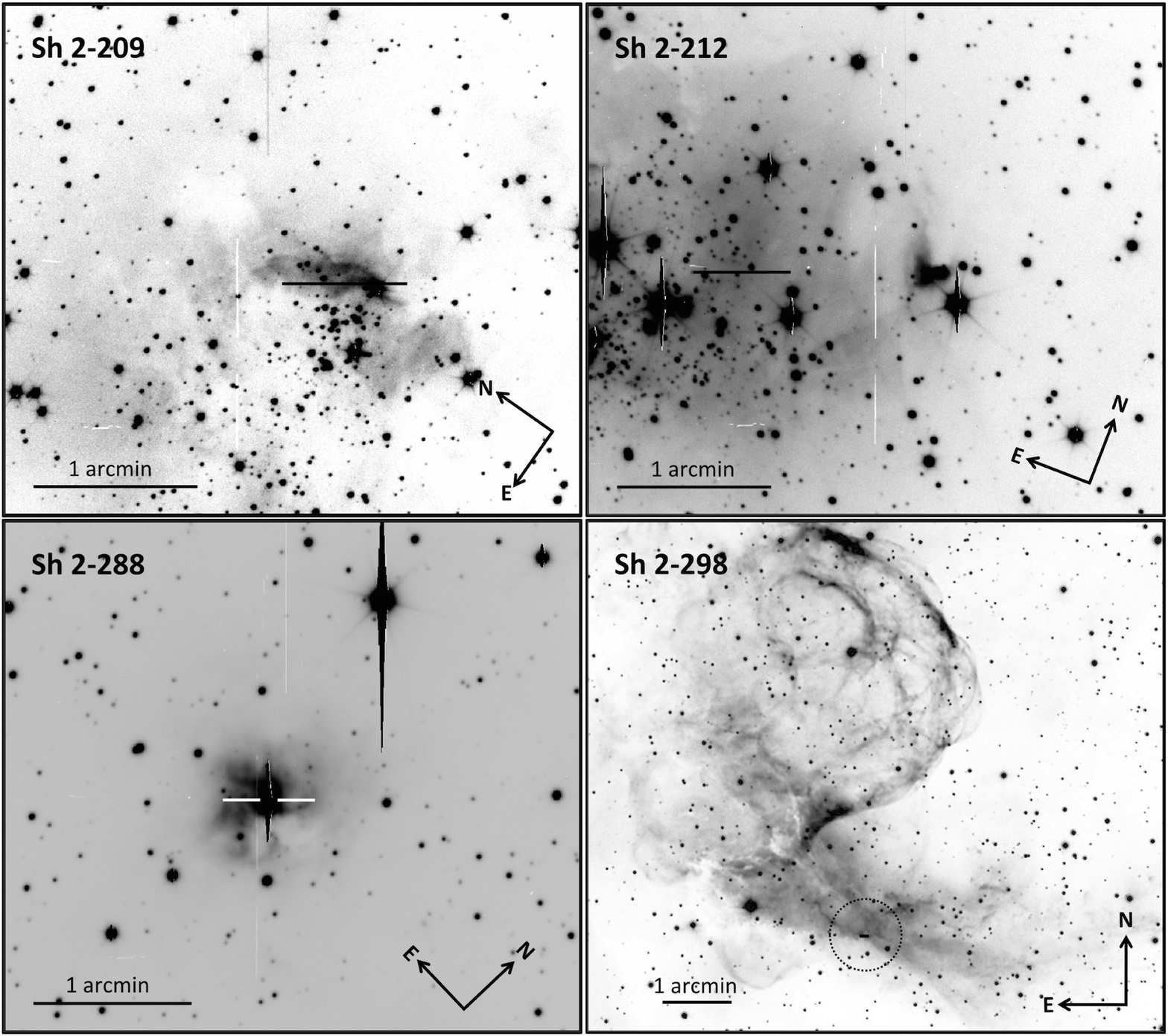} 
   \caption{
  The GTC {\it g}-band acquisition images of Sh~2-209, Sh~2-212 and Sh~2-288. The image of Sh~2-298 is 
  a combination of the $B$, $V$ and H$\alpha$ images obtained with the Wide Field Camera at the 2.5m INT (ORM) 
  by \'A. R. L\'opez-S\'anchez (AAO, CSIRO, Australia).  The position and length of the aperture extracted for each object are indicated.
   }
   \label{fig:slits_2}
  \end{figure*}

\subsection{Observations}
\label{sec:obs:1}

Except Sh~2-298, observations of all objects were performed 
with the 10.4m Gran Telescopio Canarias (GTC) at Observatorio del 
Roque de los Muchachos (ORM, La Palma, Spain). They were carried out 
in fourteen 1~hr observing blocks distributed in several nights 
between September and November in 2015.   The spectra were 
obtained with the OSIRIS (Optical System for Imaging and 
low-Intermediate-Resolution Integrated Spectroscopy) spectrograph 
\citep{cepaetal00, cepaetal03}. OSIRIS consists of a mosaic of two 
Marconi CCD42-82 CCDs (CCD1$+$CCD2), each with 2048$\times$4096 
pixels, and a 74-pixel gap between them. Each pixel has a physical 
size of 15$\mu$m.  The standard readout mode of 2$\times$2 pixel 
binning was adopted during observations, which gives a plate scale 
of 0.254~arcsec.  The long-slit spectroscopy mode was used in our 
OSIRIS observations, and all targets were located at the centre of 
CCD2. 
The slit length was 7.4 arcmin and its width was set at 0.8 arcsec, 
same as the typical seeing at ORM.  Two OSIRIS grisms, R1000B 
and R2500V, were used for the spectroscopy.  R1000B covers almost the 
whole optical wavelength range, from 3600 to 7750 \AA, while R2500V 
covers from 4430 to 6020 \AA\ but with higher spectral resolution 
for the wavelength region where the {\elect}-sensitive \foiii\ 
$\lambda$4363 line lies. 
The effective spectral resolution (full-width at half maximum, 
FWHM) was 6.52 \AA\ for the R1000B grism and 2.46 \AA\ for R2500V. 
The long slit covers the brightest regions of the nebulae. 
Coordinates of the slit centre and the position angle (PA) of 
the long-slit are given in Table~\ref{tab:journal}, where we also 
present the integration time of each object.  In the two-dimensional 
(2D) spectrum of each object, we selected the brightest region along 
the slit to extract the highest signal-to-noise 1D spectrum.  
In Figs.~\ref{fig:slits_1} and \ref{fig:slits_2} we mark the 
slit position and aperture of spectral extraction for each object. 
The aperture size of extraction along the slit is indicated in 
the seventh column of Table~\ref{tab:journal}. The OSIRIS spectra 
were wavelength calibrated using the HgAr, Ne and Xe arc lamps.

The observations of Sh~2-298 were made on November 14 and December 
2, 2003 with the Ultraviolet Visual Echelle Spectrograph 
\citep[UVES,][]{dodoricoetal00}, on the Very Large Telescope (VLT) 
Kueyen unit at Cerro Paranal Observatory (Chile).  The adopted 
standard settings in both the red and blue arms of the spectrograph 
cover a broad region from 3100 to 10\,400~\AA. The wavelength intervals 
5783--5830~\AA\ and 8540--8650~\AA\ were not observed due to a gap 
between the two CCDs used in the red arm.  There are also five 
narrow gaps that were not observed, 9608--9612~\AA, 9761--9767~\AA, 
9918--9927~\AA, 10\,080--10\,093~\AA\ and 10\,249--10\,264~\AA, 
because the five redmost orders did not fit completely within the CCD. 
The atmospheric dispersion corrector (ADC) was used to keep the same 
observed region within the slit regardless of the air mass value. 
The slit width was set to 3~arcsec and the slit length 8~arcsec. 
The effective resolution at a given wavelength is approximately 
$\Delta\lambda \sim \lambda / 8800$. 
Spectroscopic observations of the spectrophotometric standard 
star HD~49798 \citep{turnsheketal90,bohlinlindler92} were made 
for flux calibration.

\subsection{Data Reduction}
\label{sec:obs:2}

The GTC OSIRIS long-slit spectra were reduced using 
{\sc iraf}\footnote{{\sc iraf}, the Image Reduction and Analysis 
Facility, is distributed by the National Optical Astronomy Observatory, 
which is operated by the Association of Universities for Research 
in Astronomy under cooperative agreement with the National Science 
Foundation.} v2.16. 
Data reduction followed the standard procedure for long-slit 
spectra.  We first removed cosmic rays by combining the raw 
spectrograms of each {\hii} region, and then subtracted the bias 
and corrected for the flat field.  We then carried out wavelength 
calibration using the arc-line spectra.  According to the coverage 
of arc lines across the wavelength range of the grism, we used the 
HgAr arc lines for calibration of the R1000B spectra, and 
HgAr+Ne+Xe for the R2500V spectra.  Geometry distortion of emission 
lines along the long slit exists for extended sources.  During the 
wavelength calibration, this geometry distortion was also rectified 
by fitting the arc lines with sixth- or seventh-order polynomial 
functions in the 2D spectrogram.  We then applied the resolution 
derived from fitting of the arc-line spectrogram to the target 
spectrum.  Through this procedure, all nebular emission lines (also 
the sky emission lines) in a 2D spectrum were ``straightened'' 
along the long slit. 

Particular care was taken in background subtraction because the sky 
background emission is inhomogeneous along the GTC OSIRIS long slit 
\citep{fangetal15}.  The emission profiles of sky background 
along the slit direction was first fitted with polynomial functions, 
and then subtracted from the spectrogram.  Since our targets are all 
extended sources, low-order polynomial functions were adopted for 
the profile fitting so that the true nebular emission was not removed 
due to possible over-subtraction.  The {\sc background} package in 
{\sc iraf} was used in the background subtraction.  This procedure 
produced a cleaned spectrogram for each target, which was then 
flux-calibrated (and also corrected for the atmospheric extinction) 
using the spectrum of a spectrophotometric standard star.

  \begin{figure*}
   \centering
   \includegraphics[scale=1.0]{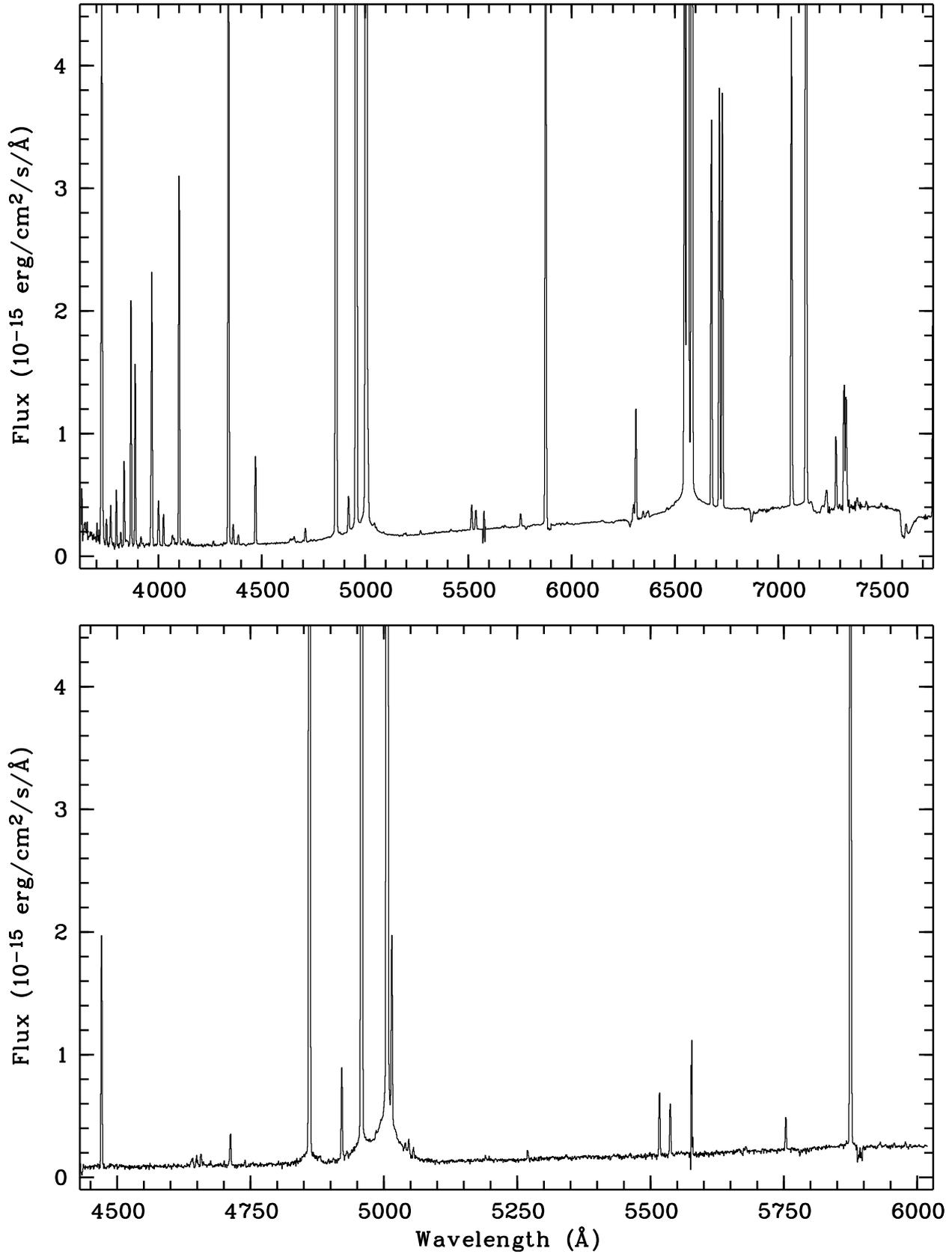} 
   \caption{
   The GTC OSIRIS long-slit 1D spectra of Sh~2-100 obtained 
   with the R1000B (top panel) and R2500V (bottom panel) grisms. 
   Spectra have been fully calibrated, but interstellar extinction 
   has not been corrected for.  
   }
   \label{fig:sh2-100}
  \end{figure*}

Finally, we extracted 1D spectrum from the cleaned, fully calibrated 
2D frame for each {\hii} region, using the slit aperture indicated in 
Figs~\ref{fig:slits_1} and \ref{fig:slits_2}.  As an example, 
Fig.~\ref{fig:sh2-100} shows the extracted 1D spectra for Sh~2-100. 
The temperature-sensitive [O~{\sc iii}] $\lambda$4363 and [N~{\sc ii}] 
$\lambda$5755 auroral lines are well detected.  In the sections below, 
we will also analyse the optical recombination lines of C~{\sc ii} and 
O~{\sc ii} that were also observed in the deep spectra of this target. 
It is worth mentioning that the second-order contamination exists in 
the red part ($>$6300 \AA) of the OSIRIS R1000B grism.  We corrected 
for this effect by fitting the overall shape of the derived R1000B 
efficiency curve with a fifth-order polynomial function.  This 
correction has proved to be reliable \citep{fangetal15}.

The VLT UVES echelle spectra of Sh~2-298 were reduced using the 
{\sc echelle} reduction package in {\sc iraf}, following the standard 
procedure of bias subtraction, aperture extraction, flat-fielding, 
wavelength calibration and then flux calibration.

\section{Line intensity measurements}
\label{sec:lines}

Emission line fluxes of the spectra of the 8 {\hii} regions included 
in Table~\ref{tab:journal} were measured with the {\sc splot} routine 
of {\sc iraf} by integrating over the emission line profile between 
two given limits and over the average local continuum.  All line 
fluxes of a given spectrum have been normalized to H$\beta$ = 100. 
In the wavelength range between 4430 and 6020\,\AA\ of the GTC 
spectra, the emission line fluxes measured from the R2500V spectrum 
were used instead of those measured from the R1000B spectrum, although 
our flux calibration was reliable and the spectral-line measurements 
obtained with the two OSIRIS grisms agree well with each other. 
The S/Ns of faint emission lines are higher in the spectra taken with 
the grism R2500V.  In the case of line blending, we applied double- 
or multiple-Gaussian profile fit using the {\sc splot} routine of 
{\sc iraf} to measure the individual line intensities. 
Identification of emission lines were made following our previous 
works on spectroscopy of bright Galactic {\hii} regions 
\citep[see][and references therein]{garciarojasesteban07}. 

  \setcounter{table}{1}  
  \begin{table*}
  \centering
     \caption{Derredened line intensity ratios with respect to $I$(\hb) = 100 for Sh~2-100, Sh~2-128, Sh~2-288 and Sh~2-127.}
     \label{tab:lines_1}
    \begin{tabular}{lccccccc}
     \hline                                              
        $\lambda_0$ & & & & \multicolumn{4}{c}{$I$($\lambda$)/$I$(\hb)} \\  
       (\AA) & Ion & ID& f($\lambda$) & Sh 2-100 & Sh 2-128 & Sh 2-288 & Sh 2-127 \\ 
     \hline  
3704	&	{\hi}	&	H16	&	0.260	&	0.99	$\pm$	0.22	&		$-$		&		$-$		&		$-$		\\	
3705	&	{\hei}	&	25	&		&				&				&				&				\\	
3712	&	{\hi}	&	H15	&	0.259	&	0.67	$\pm$	0.23	&		$-$		&		$-$		&		$-$		\\	
3726	&	{\foii}	&	1F	&	0.257	&	68.0	$\pm$	4.4	&	217.5	$\pm$	6.9	&	317	$\pm$	22	&	366	$\pm$	11	\\	
3729	&	{\foii}	&	1F	&		&				&				&				&				\\	
3750	&	{\hi}	&	H12	&	0.253	&	2.25	$\pm$	0.26	&		$-$		&	3.21	$\pm$	0.45	&	8.2:		\\	
3771	&	{\hi}	&	H11	&	0.249	&	3.36	$\pm$	0.44	&		$-$		&	2.92	$\pm$	0.41	&		$-$		\\	
3798	&	{\hi}	&	H10	&	0.244	&	4.53	$\pm$	0.44	&	5.5	$\pm$	2.0	&	3.73	$\pm$	0.68	&		$-$		\\	
3820	&	{\hei}	&	22	&	0.240	&	1.02	$\pm$	0.17	&		$-$		&		$-$		&		$-$		\\	
3835	&	{\hi}	&	H9	&	0.237	&	7.00	$\pm$	0.53	&	7.9	$\pm$	1.5	&	6.31	$\pm$	0.69	&	10.2	$\pm$	2.5	\\	
3867	&	{\hei}	&	20	&	0.231	&	20.6	$\pm$	1.3	&	9.8	$\pm$	1.8	&		$-$		&		$-$		\\	
3869	&	{\fneiii} 	&	1F	&		&				&				&				&				\\	
3889	&	{\hei}	&	5	&	0.227	&	14.93	$\pm$	0.90	&	17.1	$\pm$	1.4	&	16.3	$\pm$	1.2	&	20.4	$\pm$	2.9	\\	
3889	&	{\hi}	&	H8	&		&				&				&				&				\\	
3967	&	{\fneiii} 	&	1F	&	0.211	&	23.3	$\pm$	1.3	&	20.3	$\pm$	1.3	&		$-$		&		$-$		\\	
3970	&	{\hi}	&	H7	&		&				&				&	15.7	$\pm$	1.1	&		$-$		\\	
4026	&	{\hei}	&	18	&	0.198	&	2.22	$\pm$	0.17	&		$-$		&	1.68	$\pm$	0.18	&		$-$		\\	
4069	&	{\fsii}	&	1F	&	0.189	&	0.727	$\pm$	0.051	&		$-$		&	1.40	$\pm$	0.18	&		$-$		\\	
4076	&	{\fsii}	&	1F	&	0.187	&	0.339	$\pm$	0.024	&		$-$		&	0.37	$\pm$	0.14	&		$-$		\\	
4102	&	{\hi}	&	H6	&	0.182	&	25.96	$\pm$	1.23	&	26.2	$\pm$	1.2	&	26.1	$\pm$	1.3	&	26.1	$\pm$	1.3	\\	
4144	&	{\hei}	&	53	&	0.172	&	0.293	$\pm$	0.069	&		$-$		&		$-$		&		$-$		\\	
4156	&	{\nii}	&	19	&	0.171	&	0.125	$\pm$	0.043	&		$-$		&		$-$		&		$-$		\\	
4267	&	{\cii}	&	6	&	0.144	&	0.249	$\pm$	0.049	&		$-$		&		$-$		&		$-$		\\	
4340	&	{\hi}	&	H$\gamma$	&	0.127	&	44.44	$\pm$	1.60	&	45.0	$\pm$	1.0	&	44.4	$\pm$	1.7	&	44.8	$\pm$	1.1	\\	
4363	&	{\foiii}	&	2F	&	0.121	&	1.145	$\pm$	0.065	&	1.76	$\pm$	0.18	&	0.58	$\pm$	0.11	&		$-$		\\	
4388	&	{\hei}	&	51	&	0.115	&	0.529	$\pm$	0.046	&		$-$		&		$-$		&		$-$		\\	
4471	&	{\hei}	&	14	&	0.096	&	4.85	$\pm$	0.15	&	4.18	$\pm$	0.23	&	3.00	$\pm$	0.15	&	2.77	$\pm$	0.25	\\	
4607	&	{\ffeiii}	&	3F	&	0.062	&		$-$		&		$-$		&	0.076	$\pm$	0.020	&		$-$		\\	
4639	&	{\oii}	&	1	&	0.055	&	0.217	$\pm$	0.038	&		$-$		&	0.062	$\pm$	0.023	&		$-$		\\	
4642	&	{\oii}	&	1	&		&				&				&				&				\\	
4643	&	{\nii}	&	5	&		&				&				&				&				\\	
4649	&	{\oii}	&	1	&	0.052	&	0.183	$\pm$	0.029	&		$-$		&		$-$		&		$-$		\\	
4651	&	{\oii}	&	1	&		&				&				&				&				\\	
4658	&	{\ffeiii}	&	3F	&	0.050	&	0.213	$\pm$	0.019	&	0.300	$\pm$	0.046	&	1.086	$\pm$	0.045	&	0.645	$\pm$	0.052	\\	
4662	&	{\oii}	&	1	&	0.049	&	0.041	$\pm$	0.008	&		$-$		&		$-$		&		$-$		\\	
4676	&	{\oii}	&	1	&	0.043	&	0.036	$\pm$	0.009	&		$-$		&		$-$		&		$-$		\\	
4702	&	{\ffeiii}	&	3F	&	0.039	&	0.072	$\pm$	0.011	&		$-$		&	0.262	$\pm$	0.039	&		$-$		\\	
4711	&	{\fariv}	&	1F	&	0.037	&	0.598	$\pm$	0.035	&	0.516	$\pm$	0.065	&	0.303	$\pm$	0.029	&	0.500	$\pm$	0.082	\\	
4713	&	{\hei}	&	12	&	0.036	&				&				&				&				\\	
4734	&	{\ffeiii}	&	3F	&	0.031	&		$-$		&		$-$		&	0.075	$\pm$	0.013	&		$-$		\\	
4740	&	{\fariv}	&	1F	&	0.030	&	0.090	$\pm$	0.019	&		$-$		&		$-$		&		$-$		\\	
4755	&	{\ffeiii}	&	3F	&	0.026	&		$-$		&		$-$		&	0.198	$\pm$	0.014	&		$-$		\\	
4770	&	{\ffeiii}	&	3F	&	0.023	&		$-$		&		$-$		&	0.049	$\pm$	0.007	&		$-$		\\	
4778	&	{\ffeiii}	&	3F	&	0.021	&		$-$		&		$-$		&	0.017	$\pm$	0.008	&		$-$		\\	
4789	&	{\ffeiii}	&		&	0.018	&		$-$		&		$-$		&	0.044	$\pm$	0.015	&		$-$		\\	
4815	&	{\ffeii}	&	20F	&	0.012	&		$-$		&		$-$		&	0.085	$\pm$	0.018	&		$-$		\\	
4861	&	{\hi}	&	{\hb}	&	0.000	&	100.0	$\pm$	2.0	&	100.0	$\pm$	2.0	&	100.0	$\pm$	2.0	&	100.0	$\pm$	2.0	\\	
4881	&	{\ffeiii}	&	2F	&	-0.005	&		$-$		&		$-$		&	0.237	$\pm$	0.029	&	0.222	$\pm$	0.051	\\	
4922	&	{\hei}	&	48	&	-0.015	&	1.323	$\pm$	0.074	&	1.200	$\pm$	0.088	&	0.738	$\pm$	0.041	&	0.855	$\pm$	0.051	\\	
4959	&	{\foiii}	&	1F	&	-0.024	&	126.9	$\pm$	2.6	&	90.6	$\pm$	1.8	&	41.09	$\pm$	0.87	&	20.02	$\pm$	0.41	\\	
4987	&	{\ffeiii}	&	2F	&	-0.031	&		$-$		&	0.192	$\pm$	0.039	&	0.274	$\pm$	0.022	&		$-$		\\	
5007	&	{\foiii}	&	1F	&	-0.036	&	341.9	$\pm$	7.4	&	271.0	$\pm$	5.5	&	117.0	$\pm$	2.6	&	58.7	$\pm$	1.2	\\	
5016	&	{\hei}	&	4	&	-0.038	&	2.34	$\pm$	0.30	&	2.386	$\pm$	0.072	&	1.795	$\pm$	0.061	&	1.864	$\pm$	0.060	\\	
5035	&	{\ffeii}	&	4F	&	-0.043	&		$-$		&		$-$		&	0.031	$\pm$	0.002	&		$-$		\\	
5048	&	{\hei}	&	47	&	-0.046	&	0.179	$\pm$	0.029	&	0.265	$\pm$	0.048	&	0.109	$\pm$	0.024	&		$-$		\\	
5056	&	\ion{Si}{ii}	&	5	&	-0.048	&	0.120	$\pm$	0.010	&	0.144	$\pm$	0.028	&	0.215	$\pm$	0.020	&		$-$		\\	
5159	&	{\ffeii}	&	19F	&	-0.073	&		$-$		&		$-$		&	0.067	$\pm$	0.015	&		$-$		\\	
5192	&	{\fariii}	&	3F	&	-0.081	&	0.043	$\pm$	0.007	&		$-$		&		$-$		&		$-$		\\	
5198	&	{\nitroi}	&	1F	&	-0.082	&	0.015	$\pm$	0.002	&	0.183	$\pm$	0.032	&	0.763	$\pm$	0.033	&	0.123	$\pm$	0.031	\\	
5200	&	{\nitroi}	&	1F	&	-0.083	&		$-$		&		$-$		&		$-$		&	0.055:		\\	
5262    &       {\ffeii}        &       19F     &       -0.098  &               $-$             &               $-$             &       0.072   $\pm$   0.007   &               $-$             \\
5270    &       {\ffeiii}       &       1F      &       -0.111  &       0.099   $\pm$   0.012   &       0.262   $\pm$   0.022   &       0.518   $\pm$   0.027   &       0.414   $\pm$   0.036   \\
5518    &       {\fcliii}       &       1F      &       -0.154  &       0.523   $\pm$   0.025   &       0.532   $\pm$   0.035   &       0.405   $\pm$   0.022   &       0.423   $\pm$   0.046   \\
\end{tabular}
\end{table*}
\setcounter{table}{1}
   \begin{table*}
  \centering
     \caption{continued}
    \begin{tabular}{lccccccccc}
     \hline  
        $\lambda_0$ & & & & \multicolumn{4}{c}{$I$($\lambda$)/$I$(\hb)} \\  
       (\AA) & Ion & ID& f($\lambda$) & Sh 2-100 & Sh 2-128 & Sh 2-288 & Sh 2-127 \\ 
     \hline  
5538	&	{\fcliii}	&	1F	&	-0.158	&	0.422	$\pm$	0.024	&	0.426	$\pm$	0.028	&	0.305	$\pm$	0.019	&	0.393	$\pm$	0.056	\\	
5755	&	{\fnii}	&	3F	&	-0.194	&	0.188	$\pm$	0.013	&	0.616	$\pm$	0.028	&	0.864	$\pm$	0.047	&	1.290	$\pm$	0.043	\\	
5876	&	{\hei}	&	11	&	-0.215	&	15.65	$\pm$	0.85	&	13.01	$\pm$	0.33	&	8.78	$\pm$	0.51	&	9.51	$\pm$	0.25	\\	
5958	&	{\oi}	&	23	&	-0.228	&		$-$		&		$-$		&	0.074	$\pm$	0.014	&		$-$		\\	
5979	&	\ion{Si}{ii}	&	4	&	-0.231	&		$-$		&		$-$		&	0.093	$\pm$	0.008	&		$-$		\\	
6046	&	{\oi}	&	22	&	-0.242	&		$-$		&		$-$		&	0.068	$\pm$	0.012	&		$-$		\\	
6300	&	{\foi}	&	1F	&	-0.282	&	0.157	$\pm$	0.026	&	0.833	$\pm$	0.041	&	1.10	$\pm$	0.11	&		$-$		\\	
6312	&	{\fsiii}	&	3F	&	-0.283	&	1.44	$\pm$	0.11	&	1.456	$\pm$	0.060	&	1.29	$\pm$	0.11	&	1.83	$\pm$	0.13	\\	
6347	&	\ion{Si}{ii}	&	4	&	-0.289	&	0.082	$\pm$	0.017	&	0.040	$\pm$	0.010	&	0.139	$\pm$	0.026	&	0.116	$\pm$	0.025	\\	
6364	&	{\oi}	&	1F	&	-0.291	&	0.047	$\pm$	0.012	&	0.278	$\pm$	0.031	&	0.493	$\pm$	0.082	&		$-$		\\	
6371	&	\ion{Si}{ii}	&	2	&	-0.292	&	0.080	$\pm$	0.019	&		$-$		&		$-$		&		$-$		\\	
6548	&	{\fnii}	&	1F	&	-0.318	&	5.77	$\pm$	0.45	&	11.21	$\pm$	0.33	&	21.6	$\pm$	1.8	&	30.01	$\pm$	0.89	\\	
6563	&	{\hi}	&	{\ha}	&	-0.320	&	100.1	$\pm$	7.8$^{\rm a}$	&	288.7	$\pm$	8.6	&	183	$\pm$	15$^{\rm a}$	&	289.1	$\pm$	8.6	\\	
6583	&	{\fnii}	&	1F	&	-0.323	&	19.5	$\pm$	1.5	&	36.8	$\pm$	1.1	&	68.0	$\pm$	5.7	&	90.1	$\pm$	2.7	\\	
6678	&	{\hei}	&	46	&	-0.336	&	4.11	$\pm$	0.33	&	3.49	$\pm$	0.12	&	2.20	$\pm$	0.20	&	2.454	$\pm$	0.083	\\	
6716	&	{\fsii}	&	2F	&	-0.342	&	4.19	$\pm$	0.35	&	8.66	$\pm$	0.27	&	13.4	$\pm$	1.2	&	13.38	$\pm$	0.41	\\	
6731	&	{\fsii}	&	2F	&	-0.344	&	4.09	$\pm$	0.34	&	8.52	$\pm$	0.27	&	13.0	$\pm$	1.2	&	14.09	$\pm$	0.44	\\	
7002	&	{\oi}	&	21	&	-0.379	&		$-$		&		$-$		&	0.058	$\pm$	0.012	&		$-$		\\	
7065	&	{\hei}	&	10	&	-0.387	&	4.39	$\pm$	0.41	&	2.98	$\pm$	0.10	&	3.22	$\pm$	0.32	&	2.478	$\pm$	0.086	\\	
7136	&	{\fariii}	&	1F	&	-0.396	&	13.7	$\pm$	1.3	&	10.77	$\pm$	0.37	&	5.79	$\pm$	0.59	&	6.48	$\pm$	0.22	\\	
7155	&	{\ffeii}	&	14F	&	-0.399	&		$-$		&		$-$		&	0.036	$\pm$	0.010	&		$-$		\\	
7236	&	{\cii}	&	3	&	-0.408	&		$-$		&		$-$		&	0.098	$\pm$	0.021	&		$-$		\\	
7281	&	{\hei}	&	45	&	-0.414	&	0.554	$\pm$	0.059	&	0.540	$\pm$	0.036	&	0.353	$\pm$	0.066	&	0.278	$\pm$	0.045	\\	
7318	&	{\foii}	&	2F	&	-0.418	&	0.82	$\pm$	0.10	&	2.71	$\pm$	0.10	&	3.92	$\pm$	0.43	&	4.67	$\pm$	0.33	\\	
7320	&	{\foii}	&	2F	&		&				&				&				&				\\	
7330	&	{\foii}	&	2F	&	-0.420	&	0.735	$\pm$	0.092	&	2.480	$\pm$	0.093	&	3.38	$\pm$	0.37	&	4.63	$\pm$	0.32	\\	
7331	&	{\foii}	&	2F	&		&				&				&				&				\\	
7751	&	{\fariii}	&	2F	&	-0.467	&		$-$		&	1.284	$\pm$	0.059	&		$-$		&	0.762	$\pm$	0.080	\\	
\multicolumn{4}{l}{$c$(H$\beta$)} & 1.73 $\pm$ 0.10 & 2.27 $\pm$ 0.03 &1.34 $\pm$ 0.11 &  2.12 $\pm$ 0.03 \\
\multicolumn{4}{l}{$F$(H$\beta$)$^{\rm b}$} & 14.25 $\pm$ 0.29 & 1.59 $\pm$ 0.03 & 9.45 $\pm$ 0.19 & 0.43 $\pm$ 0.09 \\
     \hline                                              
    \end{tabular}
    \begin{description}
      \item[$^{\rm a}$] Saturated in long exposure spectrum.  
      \item[$^{\rm b}$] Flux uncorrected for reddening in units of 10$^{-14}$ erg cm$^{-2}$ s$^{-1}$.                   
    \end{description}
   \end{table*}	
 
    
  \setcounter{table}{2}  
  \begin{table*}
  \centering
     \caption{Derredened line intensity ratios with respect to $I$(\hb) = 100 for Sh~2-212, Sh~2-83 and Sh~2-209.}
     \label{tab:lines_2}
    \begin{tabular}{lcccccc}
     \hline                                              
        $\lambda_0$ & & & & \multicolumn{3}{c}{$I$($\lambda$)/$I$(\hb)} \\  
       (\AA) & Ion & ID& f($\lambda$) & Sh 2-212 & Sh 2-83 & Sh 2-209 \\ 
     \hline  
3726	&	{\foii}	&	1F	&	0.257	&	170.9	$\pm$	5.9	&	76.1	$\pm$	5.7	&		$-$		\\							
3729	&	{\foii}	&	1F	&		&				&				&				\\							
3750	&	{\hi}	&	H12	&	0.253	&	6.5	$\pm$	1.4	&		$-$		&		$-$		\\							
3771	&	{\hi}	&	H11	&	0.249	&	9.7	$\pm$	2.2	&		$-$		&		$-$		\\							
3835	&	{\hi}	&	H9	&	0.237	&	6.1	$\pm$	1.9	&		$-$		&	97	$\pm$	22	\\							
3867	&	{\hei}	&	20	&	0.231	&	9.1	$\pm$	2.7	&	44.2	$\pm$	4.3	&		$-$		\\							
3869	&	{\fneiii} 	&	1F	&		&				&				&				\\							
3889	&	{\hei}	&	5	&	0.227	&	18.4	$\pm$	2.1	&	14.2	$\pm$	3.0	&		$-$		\\							
3889	&	{\hi}	&	H8	&		&				&				&				\\							
3967	&	{\fneiii} 	&	1F	&	0.211	&	16.7	$\pm$	2.4	&	17.8	$\pm$	2.4	&		$-$		\\							
3970	&	{\hi}	&	H7	&		&				&				&				\\							
4102	&	{\hi}	&	H6	&	0.182	&	25.0	$\pm$	2.1	&	23.8	$\pm$	1.7	&		$-$		\\							
4340	&	{\hi}	&	H$\gamma$	&	0.127	&	33.9	$\pm$	1.2	&	45.3	$\pm$	1.3	&	39.5	$\pm$	2.8	\\							
4363	&	{\foiii}	&	2F	&	0.121	&	2.48	$\pm$	0.61	&	4.16	$\pm$	0.50	&	17.7	$\pm$	3.0	\\							
4471	&	{\hei}	&	14	&	0.096	&	4.61	$\pm$	0.25	&	4.00	$\pm$	0.36	&		$-$		\\							
4607	&	{\ffeiii}	&	3F	&	0.062	&	0.221	$\pm$	0.052	&		$-$		&		$-$		\\							
4658	&	{\ffeiii}	&	3F	&	0.050	&	0.243	$\pm$	0.069	&		$-$		&		$-$		\\							
4711	&	{\fariv}	&	1F	&	0.037	&	0.46	$\pm$	0.12	&		$-$		&		$-$		\\							
4713	&	{\hei}	&	12	&	0.036	&				&				&				\\							
4861	&	{\hi}	&	{\hb}	&	0.000	&	100.0	$\pm$	2.1	&	100.0	$\pm$	2.1	&	100.0	$\pm$	2.3	\\							
4922	&	{\hei}	&	48	&	-0.015	&	1.433	$\pm$	0.064	&	1.052	$\pm$	0.088	&		$-$		\\							
4959	&	{\foiii}	&	1F	&	-0.024	&	88.8	$\pm$	1.8	&	189.0	$\pm$	3.8	&	92.4	$\pm$	2.0	\\							
5007	&	{\foiii}	&	1F	&	-0.036	&	265.4	$\pm$	5.3	&	565	$\pm$	12	&	276.5	$\pm$	6.1	\\							
5016	&	{\hei}	&	4	&	-0.038	&	2.68	$\pm$	0.24	&	2.23	$\pm$	0.25	&	1.50	$\pm$	0.27	\\							
5048	&	{\hei}	&	47	&	-0.046	&				&	0.206	$\pm$	0.037	&		$-$		\\							
5518	&	{\fcliii}	&	1F	&	-0.154	&	0.334	$\pm$	0.065	&	0.610	$\pm$	0.044	&	0.65	$\pm$	0.15	\\							
5538	&	{\fcliii}	&	1F	&	-0.158	&	0.337	$\pm$	0.058	&	0.420	$\pm$	0.031	&	0.353	$\pm$	0.085	\\							
5755	&	{\fnii}	&	3F	&	-0.194	&	0.103	$\pm$	0.026	&	0.353	$\pm$	0.034	&	0.530	$\pm$	0.091	\\							
5876	&	{\hei}	&	11	&	-0.215	&	14.04	$\pm$	0.38	&	12.93	$\pm$	0.40	&	11.55	$\pm$	0.70	\\							
6300	&	{\foi}	&	1F	&	-0.282	&		$-$		&	0.791	$\pm$	0.098	&		$-$		\\							
6312	&	{\fsiii}	&	3F	&	-0.283	&	0.605	$\pm$	0.063	&	1.566	$\pm$	0.085	&	1.11	$\pm$	0.15	\\							
6364	&	{\oi}	&	1F	&	-0.291	&		$-$		&	0.144	$\pm$	0.037	&		$-$		\\							
6548	&	{\fnii}	&	1F	&	-0.318	&	2.424	$\pm$	0.062	&	4.76	$\pm$	0.19	&	10.30	$\pm$	0.87	\\							
6563	&	{\hi}	&	{\ha}	&	-0.320	&	280.0	$\pm$	7.2	&	274	$\pm$	11	&	259	$\pm$	22	\\							
6583	&	{\fnii}	&	1F	&	-0.323	&	11.64	$\pm$	0.30	&	16.08	$\pm$	0.63	&	31.0	$\pm$	2.7	\\							
6678	&	{\hei}	&	46	&	-0.336	&	2.22	$\pm$	0.25	&	3.12	$\pm$	0.15	&	2.98	$\pm$	0.29	\\							
6716	&	{\fsii}	&	2F	&	-0.342	&	2.08	$\pm$	0.20	&	4.85	$\pm$	0.20	&	7.34	$\pm$	0.67	\\							
6731	&	{\fsii}	&	2F	&	-0.344	&	1.38	$\pm$	0.20	&	4.26	$\pm$	0.18	&	6.55	$\pm$	0.60	\\							
7065	&	{\hei}	&	10	&	-0.387	&	1.82	$\pm$	0.17	&	2.55	$\pm$	0.12	&	2.18	$\pm$	0.23	\\							
7136	&	{\fariii}	&	1F	&	-0.396	&	11.01	$\pm$	0.33	&	9.37	$\pm$	0.44	&	7.67	$\pm$	0.81	\\							
7281	&	{\hei}	&	45	&	-0.414	&		$-$		&	0.460	$\pm$	0.048	&	0.469	$\pm$	0.075	\\							
7318	&	{\foii}	&	2F	&	-0.418	&		$-$		&	1.276	$\pm$	0.070	&	1.81	$\pm$	0.30	\\							
7320	&	{\foii}	&	2F	&		&				&				&				\\							
7330	&	{\foii}	&	2F	&	-0.420	&		$-$		&	1.284	$\pm$	0.071	&	1.74	$\pm$	0.29	\\							
7331	&	{\foii}	&	2F	&		&				&				&				\\							
7751	&	{\fariii}	&	2F	&	-0.467	&		$-$		&	1.132	$\pm$	0.070	&	0.84	$\pm$	0.11	\\							
\multicolumn{4}{l}{$c$(H$\beta$)} & 0.93 $\pm$ 0.02 & 2.65 $\pm$ 0.05 & 3.32 $\pm$ 0.11 \\
\multicolumn{4}{l}{$F$(H$\beta$)$^{\rm a}$} & 4.98 $\pm$ 0.11 & 11.8 $\pm$ 0.2 & 2.93 $\pm$ 0.07 \\
     \hline                                              
    \end{tabular}
    \begin{description}
      \item[$^{\rm a}$] Flux uncorrected for reddening in units of 10$^{-15}$ erg cm$^{-2}$ s$^{-1}$.                   
    \end{description}
   \end{table*}
 

  \setcounter{table}{3}  
  \begin{table}
  \centering
     \caption{Derredened line intensity ratios with respect to $I$(\hb) = 100 for Sh 2-298.}
     \label{tab:lines_3}
    \begin{tabular}{lcccc}
     \hline                                              
        $\lambda_0$ & & &  & \\  
       (\AA) & Ion & ID& f($\lambda$) & $\lambda$)/$I$(\hb) \\ 
     \hline  
3721.83	&	{\fsiii}	&	2F	&	0.257	&	3.63	$\pm$	0.95	\\
3721.94	&	{\hi}	&	H14	&		&				\\
3726.03	&	{\foii}	&	1F	&	0.257	&	298.3	$\pm$	9.1	\\
3728.82	&	{\foii}	&	1F	&	0.256	&	417	$\pm$	12	\\
3770.63	&	{\hi}	&	H11	&	0.249	&	4.35	$\pm$	0.68	\\
3797.90	&	{\hi}	&	H10	&	0.244	&	4.22	$\pm$	0.34	\\
3819.61	&	{\hei}	&	22	&	0.240	&	1.51	$\pm$	0.18	\\
3835.39	&	{\hi}	&	H9	&	0.237	&	5.72	$\pm$	0.42	\\
3868.75	&	{\fneiii}	&	1F	&	0.230	&	96.6	$\pm$	2.7	\\
3889.05	&	{\hi}	&	H8	&	0.226	&	20.18	$\pm$	0.72	\\
3964.73	&	{\hei}	&	5	&	0.211	&	0.77	$\pm$	0.18	\\
3967.46	&	{\fneiii}	&	1F	&	0.211	&	29.52	$\pm$	0.90	\\
3970.07	&	{\hi}	&	H7	&	0.210	&	15.22	$\pm$	0.80	\\
4026.21	&	{\hei}	&	18	&	0.198	&	1.94	$\pm$	0.21	\\
4068.60	&	{\fsii}	&	1F	&	0.189	&	8.80	$\pm$	0.31	\\
4076.35	&	{\fsii}	&	1F	&	0.187	&	2.82	$\pm$	0.18	\\
4101.74	&	{\hi}	&	H6	&	0.182	&	24.77	$\pm$	0.75	\\
4340.47	&	{\hi}	&	H$\gamma$	&	0.127	&	47.5	$\pm$	1.1	\\
4363.21	&	{\foiii}	&	2F	&	0.121	&	6.36	$\pm$	0.28	\\
4387.93	&	{\hei}	&	51	&	0.115	&	0.72	$\pm$	0.15	\\
4471.48	&	{\hei}	&	14	&	0.096	&	4.59	$\pm$	0.17	\\
4562.60	&	[\ion{Mg}{i}] ?	&		&	0.073	&	0.96	$\pm$	0.08	\\
4571.09	&	\ion{Mg}{i}] ?	&		&	0.071	&	0.68	$\pm$	0.11	\\
4711.37	&	{\fariv}	&	1F	&	0.037	&	0.143	$\pm$	0.024	\\
4713.14	&	{\hei}	&	12	&	0.036	&	0.501	$\pm$	0.091	\\
4861.33	&	{\hi}	&	{\hb}	&	0.000	&	100.0	$\pm$	2.0	\\
4921.93	&	{\hei}	&	48	&	-0.015	&	1.27	$\pm$	0.10	\\
4924.5	&	{\ffeiii}	&	2F	&	-0.016	&	0.249	$\pm$	0.038	\\
4958.91	&	{\foiii}	&	1F	&	-0.024	&	199.5	$\pm$	4.0	\\
5006.84	&	{\foiii}	&	1F	&	-0.036	&	598	$\pm$	12	\\
5015.68	&	{\hei}	&	4	&	-0.038	&	1.61	$\pm$	0.27	\\
5197.90	&	{\fni}	&	1F	&	-0.082	&	1.71	$\pm$	0.29	\\
5200.26	&	{\fni}	&	1F	&	-0.083	&	2.28	$\pm$	0.34	\\
5517.71	&	{\fcliii}	&	1F	&	-0.154	&	0.92	$\pm$	0.19	\\
5537.88	&	{\fcliii}	&	1F	&	-0.158	&	0.80	$\pm$	0.22	\\
5754.64	&	{\fnii}	&	3F	&	-0.194	&	3.16	$\pm$	0.26	\\
5875.64	&	{\hei}	&	11	&	-0.215	&	12.36	$\pm$	0.45	\\
6300.30	&	{\foi}	&	1F	&	-0.282	&	27.32	$\pm$	0.83	\\
6312.10	&	{\fsiii}	&	3F	&	-0.283	&	3.61	$\pm$	0.21	\\
6347.11	&	{\silii}	&	4	&	-0.289	&	0.796	$\pm$	0.085	\\
6363.78	&	{\foi}	&	1F	&	-0.291	&	9.11	$\pm$	0.46	\\
6548.03	&	{\fnii}	&	1F	&	-0.318	&	48.9	$\pm$	1.6	\\
6562.82	&	{\hi}	&	{\ha}	&	-0.320	&	280.0	$\pm$	9.1	\\
6583.41	&	{\fnii}	&	1F	&	-0.323	&	150.7	$\pm$	4.9	\\
6678.15	&	{\hei}	&	46	&	-0.336	&	3.95	$\pm$	0.33	\\
6716.47	&	{\fsii}	&	2F	&	-0.342	&	97.6	$\pm$	3.3	\\
6730.85	&	{\fsii}	&	2F	&	-0.344	&	72.5	$\pm$	2.5	\\
7065.28	&	{\hei}	&	10	&	-0.387	&	2.179	$\pm$	0.098	\\
7135.78	&	{\fariii}	&	1F	&	-0.396	&	14.12	$\pm$	0.53	\\
7281.35	&	{\hei}	&	45	&	-0.414	&	0.318	$\pm$	0.074	\\
7318.39	&	{\foii}	&	2F	&	-0.418	&	1.62	$\pm$	0.12	\\
7319.99	&	{\foii}	&	2F	&	-0.418	&	5.03	$\pm$	0.28	\\
7329.66	&	{\foii}	&	2F	&	-0.420	&	2.78	$\pm$	0.18	\\
7330.73	&	{\foii}	&	2F	&	-0.420	&	2.57	$\pm$	0.15	\\
7751.10	&	{\fariii}	&	2F	&	-0.467	&	3.21	$\pm$	0.15	\\
8467.25	&	{\hi}	&	P17	&	-0.536	&	0.218	$\pm$	0.062	\\
8665.02	&	{\hi}	&	P13	&	-0.553	&	0.451	$\pm$	0.072	\\
8727.13	&	[\ion{C}{i}] ?	&		&	-0.558	&	0.160	$\pm$	0.021	\\
8750.47	&	{\hi}	&	P12	&	-0.560	&	0.660	$\pm$	0.051	\\
8862.79 &       {\hi}   &       P11     &       -0.569  &       0.857   $\pm$   0.081   \\
9014.91 &       {\hi}   &       P10     &       -0.581  &       1.20    $\pm$   0.11    \\
9068.90 &       {\fsiii}        &       1F      &       -0.585  &       18.39   $\pm$   0.94    \\
\end{tabular}
\end{table}
\setcounter{table}{3}
   \begin{table}
  \centering
     \caption{continued}
    \begin{tabular}{lcccc}
     \hline                                              
        $\lambda_0$ & & &  & \\  
       (\AA) & Ion & ID& f($\lambda$) & $\lambda$)/$I$(\hb) \\ 
     \hline  
9229.01	&	{\hi}	&	P9	&	-0.596	&	1.69	$\pm$	0.11	\\
9530.60	&	{\fsiii}	&	1F	&	-0.618	&	66.1	$\pm$	3.5	\\
9545.97	&	{\hi}	&	P8	&	-0.619	&	1.97	$\pm$	0.16	\\
9850.26	&	[\ion{C}{i}] ?	&		&	-0.638	&	1.83	$\pm$	0.15	\\
10049.40	&	{\hi}	&	P7	&	-0.650	&	4.05	$\pm$	0.41	\\
\multicolumn{4}{l}{$c$(H$\beta$)} & 1.12 $\pm$ 0.03 \\
\multicolumn{4}{l}{$F$(H$\beta$)$^{\rm a}$} & 2.47 $\pm$ 0.05 \\
     \hline                                              
    \end{tabular}
    \begin{description}
      \item[$^{\rm a}$] Flux uncorrected for reddening in units of 10$^{-14}$ erg cm$^{-2}$ s$^{-1}$.                   
    \end{description}
   \end{table}		
 
  
The logarithmic reddening coefficient, $c$({\hb}), was derived by 
comparing the observed flux ratios of the brightest H~{\sc i} Balmer lines 
(H$\alpha$, H$\gamma$ and H$\delta$) with the Case~B theoretical calculations 
of \cite{storeyhummer95}.  The theoretical ratios of the H~{\sc i} lines were 
calculated for the nebular physical conditions diagnosed for each {\hii} 
region, and $T_{\rm e}$ and $N_{\rm e}$ were determined following an 
iterative process (see Section~\ref{sec:conditions} for details). We have 
used the reddening function, $f(\lambda)$, normalized to H$\beta$ derived 
by \cite{cardellietal89} and assumed $R_V$ = 3.1. 
 
In tables~\ref{tab:lines_1}, \ref{tab:lines_2} and \ref{tab:lines_3} we 
present emission line measurements of the eight {\hii} regions: the 
emission line identifications are given in the first three columns; the 
reddening function, $f(\lambda)$ is in the fourth column; the dereddened 
and normalized (with respect to H$\beta$ = 100) line intensities are presented 
in the rest columns. 
The quoted line intensity errors include the uncertainties in the flux 
measurements and error propagation of the reddening coefficient. 
The reddening coefficient $c$({\hb}) and the observed H$\beta$ line 
flux $F$(H$\beta$), as measured from the extracted 1D spectrum, of each 
{\hii} region are presented in the final two rows of each table.  In the 
case of Sh~2-298 (Table~\ref{tab:lines_3}), we give two decimals for the 
laboratory wavelength of the lines because of the much higher spectral 
resolution of the VLT UVES observations.

\section{The additional sample}
\label{sec:additional}

In addition to the Galactic {\hii} regions observed in this paper, we 
have included the deep spectra of other Galactic objects available 
from the literature.  Most of these nebulae are located at $R_{\rm G}$ 
$<$11.5~kpc and only NGC~2579 lies beyond that distance.  These objects 
were selected to compare the results of the anticentre {\hii} regions 
with those in the inner Galactic disc.  Table~\ref{tab:additional} gives 
the most common designation of the additional sample of objects, their 
Galactocentric distances (see Section~\ref{sec:distances}), their O 
abundances and the references of their emission-line ratios used for 
analysis in this paper.  The data for M20, M16, M17, M8, NGC~3576, M42, 
NGC~3603, Sh~2-311 and NGC~2579 were observed with the UVES spectrograph 
at the VLT, the same instrument configuration as the observations of 
Sh~2-298 included in this paper.  The observations of NGC~7635 were 
obtained with the same instrument configuration as our observations 
obtained with the GTC OSIRIS spectrograph. The observations of IC~5146,
Sh~2-132 and Sh~2-156 produced deep optical spectra obtained with the 
ISIS spectrograph at the 4.2m WHT.  Sh~2-132 and Sh~2-156 have been 
selected from the sample of Galactic anticentre {\hii} regions 
observed by \cite{fernandezmartinetal16}.  We have not considered the 
rest of the objects observed by those authors because 
observations of three of them (Sh~2-83, Sh~2-212 and NGC~7635) 
were also reported by us in this or previous papers, and the rest do 
not show auroral lines for determination of {\elect} in their spectra. 
The emission line ratios of IC~5146 were retrieved from 
\cite{garciarojasetal14}.

\begin{table*}
   \centering
   \caption{Additional sample of {\hii} regions.}
   \label{tab:additional}
   \begin{tabular}{lccc}
        \hline
        & $R_\mathrm{G}^{\rm a}$ & & \\
        {\hii} region & (kpc) & 12 + log(O/H) &  Reference \\
        \hline                                                                                                                                                                  
         M20 & 5.1 $\pm$ 0.3 & 8.51 $\pm$ 0.04 &  \cite{garciarojasetal06} \\
         M16 & 5.9 $\pm$ 0.2 & 8.54 $\pm$ 0.04 &   \cite{garciarojasetal09} \\
         M17 & 6.1 $\pm$ 0.2 & 8.54 $\pm$ 0.04 &   \cite{garciarojasetal07} \\
         M8 & 6.3 $\pm$ 0.8 & 8.45 $\pm$ 0.04 &  \cite{garciarojasetal07} \\
         NGC~3576 & 7.5 $\pm$ 0.3 & 8.55 $\pm$ 0.04 &  \cite{garciarojasetal04} \\
         IC~5146 & 8.10 $\pm$ 0.02 & 8.56 $\pm$ 0.04 &  \cite{garciarojasetal14} \\
         M42 & 8.34 $\pm$ 0.02 & 8.50 $\pm$ 0.04 &  \cite{estebanetal04} \\  
         NGC~3603 & 8.6 $\pm$ 0.4 & 8.44 $\pm$ 0.03 &   \cite{garciarojasetal06} \\    
         Sh~2-132 & 10.0 $\pm$ 0.7 & 8.35 $\pm$ 0.14 &  \cite{fernandezmartinetal16} \\
         NGC~7635 & 10.2 $\pm$ 0.7 & 8.40 $\pm$ 0.08 & \cite{estebanetal16} \\ 
         Sh~2-156 & 10.6 $\pm$ 0.6 & 8.32 $\pm$ 0.10 & \cite{fernandezmartinetal16}  \\     
         Sh~2-311 & 11.1 $\pm$ 0.4 & 8.39 $\pm$ 0.01 &  \cite{garciarojasetal05} \\
         NGC~2579 & 12.4 $\pm$ 0.5 & 8.26 $\pm$ 0.03 & \cite{estebanetal13}  \\                
        \hline
   \end{tabular}
    \begin{description}
      \item[$^{\rm a}$] Galactocentric distances assuming the Sun is at 8~kpc.
    \end{description}
\end{table*}

\section{Distances}
\label{sec:distances}

In the studies of Galactic abundance gradients, the distances to 
{\hii} regions are highly uncertain.  However these uncertainties are 
usually not taken into account when calculating the radial gradients. 
The adopted galactocentric distances, $R_{\rm G}$, of the objects are 
presented in Tables~\ref{tab:journal} and \ref{tab:additional}.  These 
distances are estimated as the mean values of kinematic and stellar 
distances given in different published references.  Their associated 
uncertainties correspond to the standard deviation of the values 
considered for the average.  In our distance calculations, we 
assumed that the Sun is located at $R_{\rm G}$ = 8.0~kpc \citep{reid93}. 
To derive the mean value and standard deviation of $R_{\rm G}$ for each 
object, we used the kinematic distances determined by \cite{quirezaetal06} 
and \cite{balseretal11}, the stellar ones calculated by \cite{fosterbrunt15}, 
and the kinematic and stellar distances calculated or compiled by 
\cite{russeil03} and \cite{caplanetal00}.  In addition, we have also 
considered other calculations of the average distance values of NGC~2579, 
M20, NGC~3576, NGC~3603 and IC~5146.  In the case of NGC~2579, we have 
included the consistent stellar and kinematic distances obtained by 
\cite{copettietal07}.  For M20 we adopted the distance derived from a 
detailed 3D extinction map by \cite{cambresyetal11}.  For IC~5146, 
\cite{garciarojasetal14} obtained an accurate stellar distance based on 
their spectroscopic analysis of the ionising star of the nebula.  For 
NGC~3576 and NGC~3603 we have included the kinematic distances determined 
by \cite{depreeetal99}.  In general, these additional distance 
determinations are fairly consistent with the stellar and kinematic 
distances given by \cite{russeil03} for each object.  Finally, for M42 
we adopted the distance obtained from trigonometric parallax 
\citep{mentenetal07}.  We have not considered other sources of distances 
for this last object.  From Tables~\ref{tab:journal} and \ref{tab:additional}, 
we can see that the whole sample includes 21 {\hii} regions with direct 
determinations of {\elect} that covers a range of 
Galactocentric distances from 5.1 to 17~kpc, 8 objects with $R_{\rm G}$ 
$>\,R_{25}$, and 13 with $R_{\rm G}$ $<\,R_{25}$.

\section{Results}
\label{sec:results}

For the 8 objects in the observed sample, we carried out plasma 
diagnostics to determine the physical conditions ({\elect}, {\elecd}) 
and the ionic abundances based on the line-intensity ratios given in 
Tables~\ref{tab:lines_1}, \ref{tab:lines_2} and \ref{tab:lines_3}, using 
the program {\sc pyneb} v1.0.26 \citep{Luridianaetal15}.  The atomic 
data listed in Table~\ref{tab:atomic} are adopted in our spectral 
analysis. 
The physical conditions and the ionic and elemental abundances of oxygen 
of M20, M16, M17, M8, NGC~3576, M42, NGC~3603, Sh~2-311 and NGC~2579 
were determined by \cite{estebanetal15} using the same procedure and 
atomic data as for our current sample.  The same method was applied 
to the analysis of NGC~7635, which was published in \cite{estebanetal16}. 
For IC~5146, Sh~2-132 and Sh~2-156, we recalculated the relevant 
quantities following the same methodology as for the previous objects 
and using the line-intensity ratios (as well as uncertainties) retrieved 
from the references given in Table~\ref{tab:additional}. 

  \begin{figure*}
   \centering
   \includegraphics[scale=0.15]{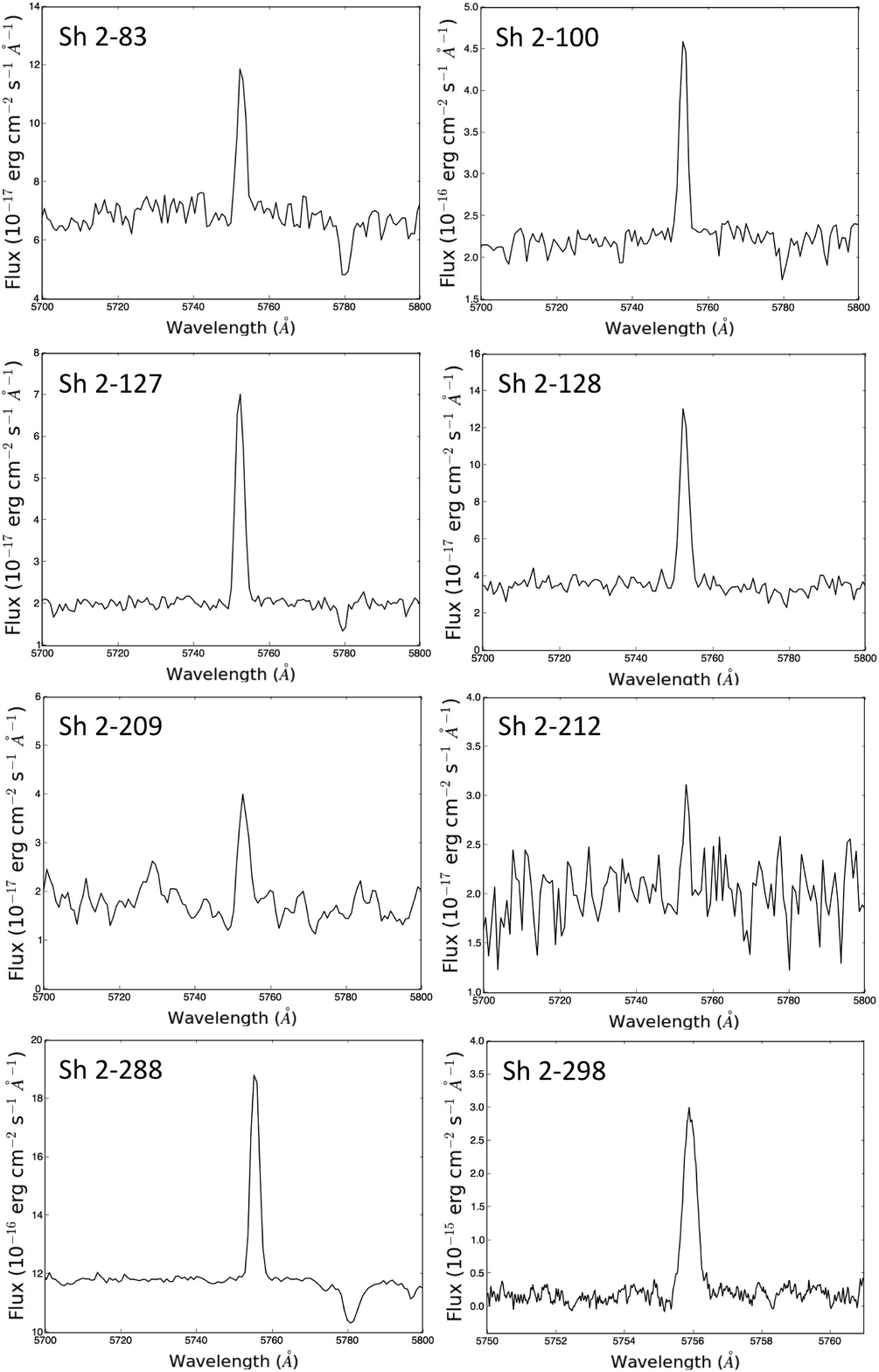} 
   \caption{Section of the spectrum of Sh~2-83, Sh~2-100, Sh~2-127, Sh~2-128,  Sh~2-209, Sh~2-212, Sh~2-288 and Sh~2-298 showing the {\fnii} $\lambda$5755 auroral line.}
   \label{fig:5755}
  \end{figure*}

  \begin{figure*}
   \centering
   \includegraphics[scale=0.15]{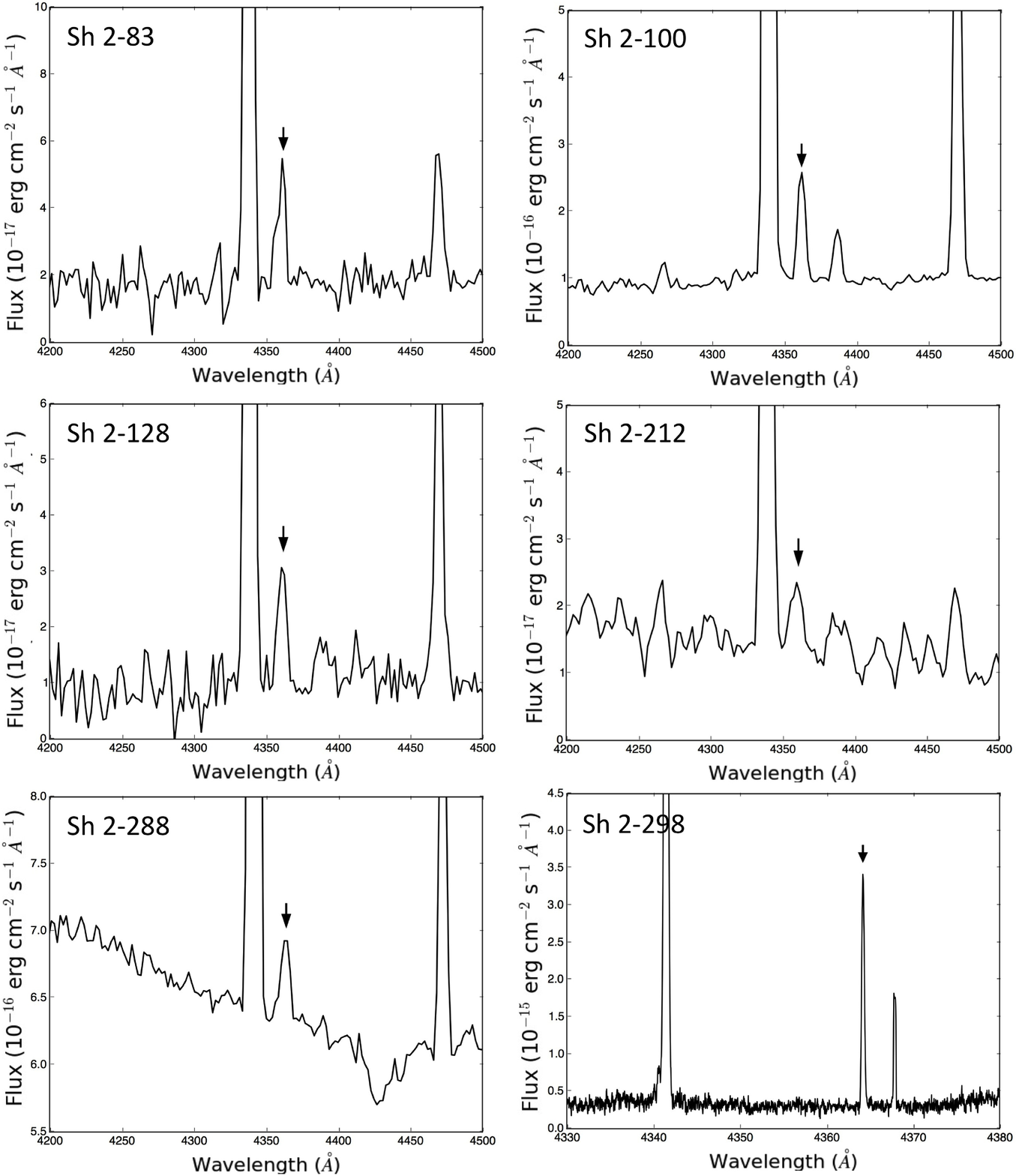} 
   \caption{Section of the spectrum of Sh~2-83, Sh~2-100, Sh~2-128, Sh~2-212, Sh~2-288 and Sh~2-298 showing the {\foiii} $\lambda$4363 auroral line (indicated by an arrow).}
   \label{fig:4363}
  \end{figure*}

  \begin{table*}
  \centering
     \caption{Atomic dataset used for collisionally excited lines.}
     \label{tab:atomic}
    \begin{tabular}{lcc}
     \hline 
	& Transition probabilities &  \\
	Ion & and energy levels & Collisional strengths \\
     \hline 
N$^+$ & \cite{froesefischertachiev04} & \cite{tayal11} \\
O$^+$ & \cite{froesefischertachiev04} & \cite{kisieliusetal09} \\
O$^{2+}$ &  \cite{froesefischertachiev04, storeyzeippen00} & \cite{storeyetal14} \\
Ne$^{2+}$ & \cite{galavisetal97} & \cite{mclaughlinbell00} \\
S$^+$ & \cite{podobedovaetal09} & \cite{tayalzatsarinny10} \\
S$^{2+}$ &  \cite{podobedovaetal09} & \cite{tayalgupta99} \\
Cl$^{2+}$ & \cite{mendoza83} & \cite{butlerzeippen89} \\
Ar$^{2+}$ & \cite{mendoza83, kaufmansugar86} & \cite{galavisetal95} \\
Ar$^{3+}$ & \cite{mendozazeippen82a, kaufmansugar86} & \cite{zeippenetal87} \\
Fe$^{2+}$ &  \cite{quinet96, johanssonetal00} & \cite{zhang96} \\
     \hline
    \end{tabular}
  \end{table*}

\subsection{Physical Conditions} 
\label{sec:conditions}

We have adopted the {\fsii}~$\lambda$6717/$\lambda$6731 and 
{\fcliii}~$\lambda$5518/$\lambda$5538 density-diagnostic line ratios for 
all the objects, and also added the {\foii}~$\lambda$3726/$\lambda$3729 
ratio for Sh~2-298.  The electron density, {\elecd}, derived from the [S~{\sc ii}] 
line ratio, {\elecd}({\fsii}), was assumed for all the apertures and 
objects, and it is alway below 1000~{\cmc} (see Tables~\ref{cond_abun_1} 
and \ref{cond_abun_2}).  For the singly ionised species, we derived 
{\elect} using the {\fnii}~($\lambda$6548+$\lambda$6584)/$\lambda$5755 
nebular-to-auroral line ratio for all objects (see Fig.~\ref{fig:5755}) 
and the {\foii}~($\lambda$7319+$\lambda$7330)/($\lambda$3726+$\lambda$3729) 
and {\fsii}~($\lambda$4068+$\lambda$4076)/($\lambda$6717+$\lambda$6731) 
ratios for some of them. 
Intensities of the {\foii} $\lambda\lambda$7319,7330 and {\fnii} 
$\lambda$5755 auroral lines have been corrected for the possible 
contribution due to recombination process using the formulae derived by 
\citet{liuetal00}.  
This contribution is between 0.2 and 13\% in the case of the {\foii} lines 
and between 0.07 and 5\% for {\fnii} $\lambda$5755. 
For the doubly ionised species, we have derived {\elect} using the 
{\foiii}~($\lambda$4959+$\lambda$5007)/$\lambda$4363 line ratio for all 
objects except Sh~2-127 and Sh~2-209, where the [O~{\sc iii}] $\lambda$4363 
auroral line was undetected (see Fig.~\ref{fig:4363}).  For these two 
{\hii} regions we estimated \elect({\foiii}) using \elect({\fnii}) and the 
empirical relation between the two temperatures given in the Equation~3 of 
\cite{estebanetal09}. 
The {\fsiii}~($\lambda$9069+$\lambda$9532)/$\lambda$6312 line ratio 
was also measured in Sh~2-298 because the spectrum of this object extends 
to the near-infrared region. The physical conditions of the {\hii} 
regions in our observed sample are presented in tables~\ref{cond_abun_1} 
and \ref{cond_abun_2}.

The physical conditions of M20, M16, M17, M8, NGC~3576, M42, NGC~3603, 
Sh~2-311 and NGC~2579 are presented in Table~3 of \cite{estebanetal15}, 
and the physical conditions adopted for the rest of the objects in the 
additional sample (IC~5146, Sh~2-132, NGC~7635 and Sh~2-156) are given 
in Table~\ref{cond_abun_3}. 
For NGC~7635, the mean values obtained for apertures 2, 3, 4, 5 and 
6 observed by \cite{estebanetal16} were adopted. 
For IC~5146, we present the average value from apertures 2, 3 and 4 
observed by \cite{garciarojasetal14} but recalculated it following the 
above mentioned procedure and using the atomic dataset indicated in 
Table~\ref{tab:atomic}.  \elect({\foiii}) is not quoted for IC~5146 
because the {\foiii} lines were not well detected in its spectra and its 
\ionic{O}{2+}/\ionic{H}{+} ratio was negligible.  In the case of Sh~2-132 
and Sh~2-156 we recalculated their {\elecd}({\fsii}), \elect({\fnii}) 
and \elect({\foiii}) using the aforementioned procedure and atomic 
dataset. The {\elect} and {\elecd} values derived for the two last 
nebulae are consistent with those obtained by \cite{fernandezmartinetal16} 
within the errors. 

  \begin{table*}
   \centering
   \caption{Physical conditions and abundances for Sh~2-100, Sh~2-128, Sh~2-288, Sh~2-127 and Sh~2-212.}
  \label{cond_abun_1}
    \begin{tabular}{l c c c c c c}
     \hline
        & \multicolumn{2}{c}{Sh~2-100} & Sh~2-128 & Sh~2-288 & Sh~2-127 & Sh~2-212 \\
     	& {\tf} = 0.0 & {\tf} = 0.010 & {\tf} = 0.0 & {\tf} = 0.0 & {\tf} = 0.0 & {\tf} = 0.0  \\
	& & $\pm$  0.010 & & & & \\
    \hline
	\multicolumn{7}{c}{Physical conditions$^{\rm a}$} \\
	\\
	{\elecd}({\fsii}) & \multicolumn{2}{c}{430 $\pm$ 210} & 480 $\pm$ 90 & 450 $\pm$ 230 & 600 $\pm$ 100 & $<$100 \\
	{\elecd}({\fcliii}) & \multicolumn{2}{c}{660 $\pm$ 350} & 700 $\pm$ 460 & 480 $\pm$ 380 & 1430 $\pm$ 1070 & 4200 $\pm$ 1900 \\
	\\
	{\elect}({\fnii}) & \multicolumn{2}{c}{8950  $\pm$  320} & 10\,970  $\pm$  260 & 9640  $\pm$  370 & 9870  $\pm$  170 & 9760  $\pm$  960 \\
	{\elect}({\foii}) & \multicolumn{2}{c}{10\,010  $\pm$  560} & 10\,040  $\pm$  220 & 9990  $\pm$  580 & 9950  $\pm$  320 & $\cdots$ \\	
	{\elect}({\fsii}) &  \multicolumn{2}{c}{19\,780  $\pm$  2450} & $\cdots$ & 9700 $\pm$  1040 &  $\cdots$ & $\cdots$ \\
	\\
	{\elect}({\foiii}) & \multicolumn{2}{c}{8140  $\pm$  120} & 9960  $\pm$  320 & 9210  $\pm$  490 & 9610  $\pm$  170$^{\rm b}$ & 10\,890  $\pm$  890 \\
	\\
	\multicolumn{7}{c}{Ionic abundances$^{\rm c}$ and O/H$^{\rm c}$ ratio} \\
	\\
	He$^+$ & \multicolumn{2}{c}{11.01 $\pm$ 0.01} & 11.00 $\pm$ 0.01 & 10.82 $\pm$ 0.01 & 10.88 $\pm$ 0.01 & 10.98 $\pm$ 0.01 \\
	C$^{2+}$ & \multicolumn{2}{c}{8.36 $\pm$ 0.09} & $\cdots$ & $\cdots$ & $\cdots$ & $\cdots$ \\
	N$^+$ & 6.71 $\pm$ 0.05 &  6.74 $\pm$  0.06 & 6.76 $\pm$  0.02 & 7.17 $\pm$  0.05 & 7.28 $\pm$  0.02 &  6.31 $\pm$ 0.09 \\ 
	O$^+$ & 7.64 $\pm$ 0.11 & 7.68 $\pm$  0.11 & 7.77 $\pm$  0.05 & 8.16 $\pm$  0.11 & 8.19 $\pm$  0.05 & 7.84 $\pm$  0.25 \\ 
	O$^{2+}$(CELs) & 8.45 $\pm$  0.02 & 8.52 $\pm$  0.06 & 7.98 $\pm$  0.04 & 7.75 $\pm$  0.07 & 7.38 $\pm$  0.03 & 7.85 $\pm$  0.09 \\ 
	O$^{2+}$(RLs) & \multicolumn{2}{c}{8.52 $\pm$ 0.06} & $\cdots$ & $\cdots$ & $\cdots$ & $\cdots$ \\
	Ne$^{2+}$ & 7.76 $\pm$  0.06 & 7.84  $\pm$  0.10 & 7.04 $\pm$ 0.14 & $\cdots$ &  $\cdots$ & 6.85 $\pm$ 0.32 \\
	S$^+$ & 5.87 $\pm$  0.07 & 5.90 $\pm$  0.07 & 5.58 $\pm$  0.03 & 5.91 $\pm$  0.11 & 5.91 $\pm$  0.02 & 4.97 $\pm$  0.12 \\ 
	S$^{2+}$ & 6.94 $\pm$  0.07 & 7.02 $\pm$  0.10 &  6.52 $\pm$  0.08 & 6.63 $\pm$  0.16 & 6.69 $\pm$  0.07 & 5.99 $\pm$  0.21 \\ 
	Cl$^{2+}$ & 5.07 $\pm$  0.03 & 5.14 $\pm$  0.08 & 4.78 $\pm$  0.05 & 4.76 $\pm$  0.08 & 4.76 $\pm$  0.06 & 4.53 $\pm$  0.13 \\ 
	Ar$^{2+}$ & 6.30 $\pm$  0.07 & 6.36 $\pm$  0.09 & 5.88 $\pm$  0.04 & 5.81 $\pm$  0.11 & 5.76 $\pm$  0.03 & 5.92 $\pm$  0.09 \\ 
	Ar$^{3+}$ & 4.83 $\pm$  0.12 & 4.90 $\pm$  0.13 & $\cdots$ & $\cdots$ & $\cdots$ & $\cdots$ \\ 	
	Fe$^{2+}$ & 5.30 $\pm$  0.04 & 5.37 $\pm$  0.09 & 5.71 $\pm$  0.06 & 5.78 $\pm$  0.05 & 5.56 $\pm$  0.04 & 4.95 $\pm$  0.28 \\
	\\
	O & 8.52 $\pm$ 0.03 & 8.58 $\pm$ 0.05 & 8.19 $\pm$ 0.03 & 8.31 $\pm$ 0.08 & 8.25 $\pm$ 0.04 & 8.15 $\pm$ 0.12 \\ 
     \hline
    \end{tabular}
    \begin{description}
      \item[$^{\rm a}$] {\elecd} in {\cmc}; {\elect} in K.
      \item[$^{\rm b}$] Estimated from {\elect}({\fnii}) and Equation~3 in \cite{estebanetal09}.       
      \item[$^{\rm c}$] In the logarithmic scale 12$+$log(\ionic{X}{n+}/\ionic{H}{+}).  
    \end{description}
  \end{table*}

  \begin{table}
   \centering
   \caption{Physical conditions and abundances for Sh~2-83, Sh~2-209 and Sh~2-298.}
  \label{cond_abun_2}
    \begin{tabular}{l c c c}
     \hline
        & Sh~2-83 & Sh~2-209 & Sh~2-298 \\
    \hline
	\multicolumn{4}{c}{Physical conditions$^{\rm a}$} \\
	\\
	{\elecd}({\fsii}) & 300 $\pm$ 100 & 310 $\pm$ 200 & $<$100 \\
	{\elecd}({\foii}) & $\cdots$ & $\cdots$ & $<$100 \\
	{\elecd}({\fcliii}) & $<$100 & $<$100 & 1100$^{+2100}_{1100}$\\
	\\
	{\elect}({\fnii}) & 12840  $\pm$  660 & 10650  $\pm$  870 & 11850  $\pm$  490 \\
	{\elect}({\foii}) & 13670  $\pm$  890 & $\cdots$ &  10060  $\pm$  200 \\	
	{\elect}({\fsii}) & $\cdots$ & $\cdots$ & 12580  $\pm$  450 \\
	\\
	{\elect}({\foiii}) & 11490  $\pm$  490 & 10710$^{\rm b}$  $\pm$  870 & 11720  $\pm$  200 \\
	{\elect}({\fsiii}) & $\cdots$ & $\cdots$ & 15340  $\pm$  700 \\
	\\
	\multicolumn{4}{c}{Ionic abundances$^{\rm c}$ and O/H$^{\rm c}$ ratio} \\
	\\
	He$^+$ & 10.94 $\pm$ 0.01 & 10.89 $\pm$ 0.02 & 10.96 $\pm$ 0.01 \\
	N$^+$ & 6.24 $\pm$ 0.04 & 6.73 $\pm$ 0.09 & 7.31 $\pm$ 0.04 \\ 
	O$^+$ & 7.05 $\pm$ 0.12 & 7.66 $\pm$ 0.23 & 8.12 $\pm$ 0.02 \\ 
	O$^{2+}$ & 8.10 $\pm$  0.05 & 7.89 $\pm$ 0.09 & 8.10 $\pm$ 0.02 \\ 
	Ne$^{2+}$ & 7.45 $\pm$  0.12 &  $\cdots$ & 7.76 $\pm$  0.03 \\
	S$^+$ & 5.14 $\pm$  0.05 & 5.50 $\pm$  0.09 & 6.49 $\pm$  0.06 \\ 
	S$^{2+}$ & 6.31 $\pm$  0.10 & 6.28 $\pm$  0.23 & 6.46 $\pm$  0.02 \\ 
	Cl$^{2+}$ & 4.63 $\pm$  0.06 & 4.70 $\pm$  0.16 & 4.83 $\pm$  0.09 \\ 
	Ar$^{2+}$ & 5.69 $\pm$  0.05 & 5.65 $\pm$  0.09 & 5.95 $\pm$  0.02\\ 
	Ar$^{3+}$ & $\cdots$ & $\cdots$ & 4.32 $\pm$  0.02  \\ 	
	\\
	O & 8.14 $\pm$ 0.05 & 8.09 $\pm$ 0.10 & 8.41 $\pm$ 0.02 \\ 
     \hline
    \end{tabular}
    \begin{description}
      \item[$^{\rm a}$] {\elecd} in {\cmc}; {\elect} in K.
      \item[$^{\rm b}$] Estimated from {\elect}({\fnii}) and Equation~3 in \cite{estebanetal09}.       
      \item[$^{\rm c}$] Assuming  {\tf} = 0.0. In the logarithmic scale 12+log(\ionic{X}{n+}/\ionic{H}{+}).  
    \end{description}
  \end{table}
 
  \begin{table*}
   \centering
   \caption{
   {\elect}, {\elecd} and abundances recalculated for IC~5146, Sh~2-132, NGC~7635 and Sh~2-156.}
  \label{cond_abun_3}
    \begin{tabular}{l c c c c}
     \hline
        & IC~5146$^{\rm a}$ & Sh~2-132 & NGC~7635$^{\rm b}$ & Sh~2-156 \\
    \hline
	{\elecd}({\fsii}) ({\cmc}) & $<$100 & 260 $\pm$ 10 & 160 $\pm$ 100 & 900 $\pm$ 25 \\
	{\elect}({\fnii}) (K) & 7140 $\pm$ 120 & 9350 $\pm$ 460 & 8390 $\pm$ 390 & 9460 $\pm$ 400 \\
	{\elect}({\foiii}) (K) & $\cdots$ & 8870$^{\rm c}$ $\pm$ 460 &  8190 $\pm$ 510 & 9010 $\pm$ 350 \\
	12+log(O/H)  & 8.56 $\pm$ 0.04 & 8.35 $\pm$ 0.14 & 8.40 $\pm$ 0.08 & 8.32 $\pm$ 0.10 \\ 
     \hline
    \end{tabular}
    \begin{description}
      \item[$^{\rm a}$] Mean of the values of apertures 2, 3 and 4 observed by \cite{garciarojasetal14}. 
      \item[$^{\rm b}$] Mean of the values of apertures 2, 3, 4, 5 and 6 observed by \cite{estebanetal16}.      
      \item[$^{\rm c}$] Estimated from \elect({\fnii}) and Equation~3 of \cite{estebanetal09}. 
    \end{description}
  \end{table*}

\subsection{Abundances} 
\label{sec:abundances}

  \begin{figure}
   \centering
   \includegraphics[scale=0.095]{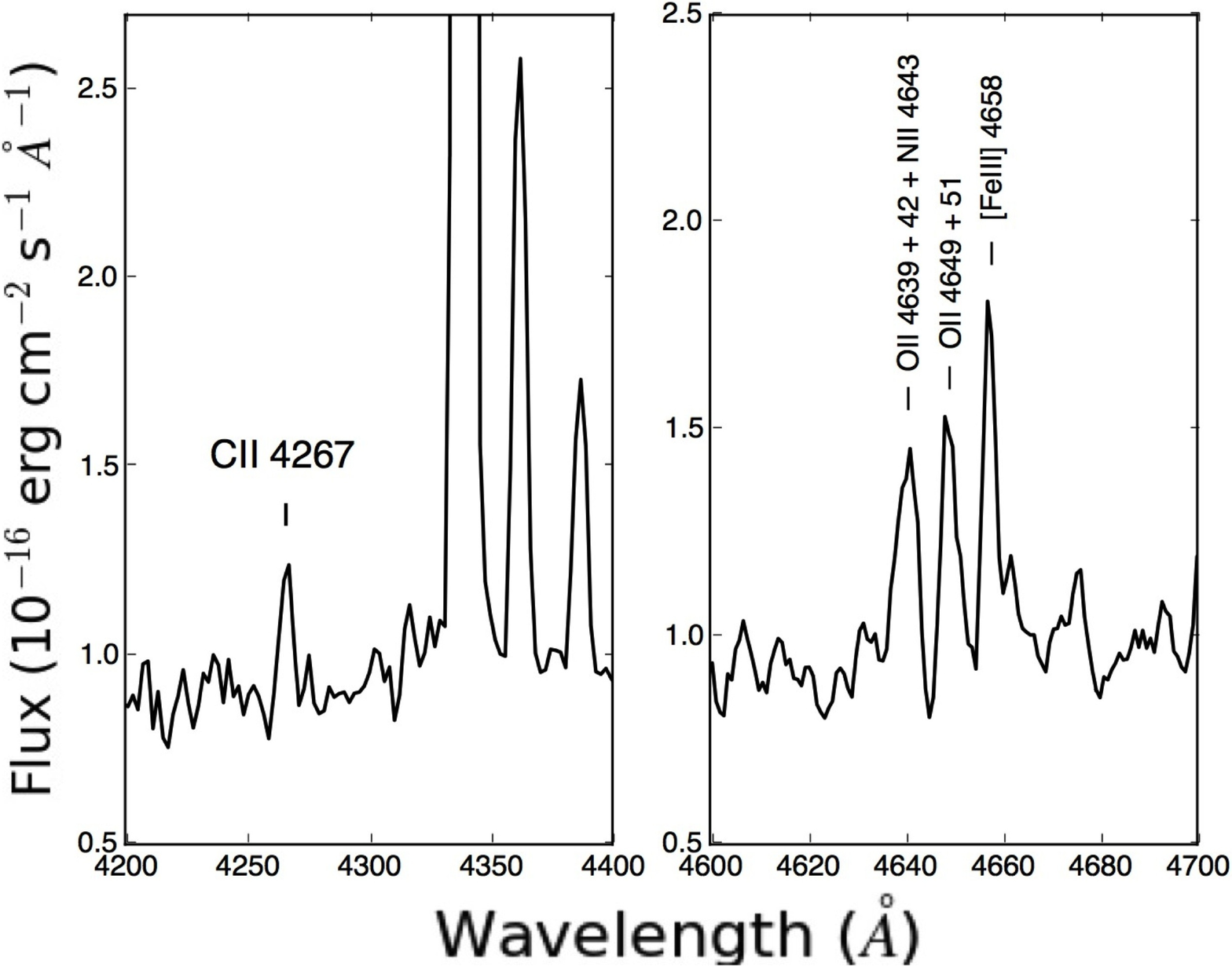} 
   \caption{Sections of the spectrum of Sh~2-100 showing the {\cii} 
   $\lambda$4267 (left) and {\oii} Multiplet 1 $\lambda$4650 (right) 
   optical recombination lines.}
   \label{fig:RLs}
  \end{figure}

The use of large-aperture telescopes, such as the GTC and VLT, allowed 
us to obtain very deep spectra where the temperature-sensitive 
faint auroral lines have been well detected in all the objects, which 
is especially important in the case of the relatively faint {\hii} 
regions in the Galactic anticentre.  With these data we have determined 
ionic abundances from the collisionally excited lines (hereafter CELs). 
In particular, we derived abundances of N$^+$, O$^+$, O$^{2+}$, S$^+$, 
S$^{2+}$, Cl$^{2+}$ and Ar$^{2+}$ for all the objects in our observed 
sample; the Ne$^{2+}$/H$^+$ ratio was derived for all nebulae except 
Sh~2-288, Sh~2-127 and Sh~2-209; the Ar$^{3+}$ abundance was derived 
only in Sh~2-100 and Sh~2-298, and the Fe$^{2+}$ abundance in Sh~2-100, 
Sh~2-128, Sh~2-288, Sh~2-127 and Sh~2-212. 
We have also determined the He$^+$/H$^+$ ratio for all the observed 
objects using the relative intensities of recombination lines (herafter 
RLs). 
We assumed a two-zone scheme for the calculations of ionic abundances. 
We adopted \elect({\fnii}) for the ions with low-ionisation potentials 
(N$^+$, O$^+$, S$^+$ and Fe$^{2+}$) and \elect({\foiii}) for the 
high-ionisation species (He$^+$, O$^{2+}$, Ne$^{2+}$, S$^{2+}$, Cl$^{2+}$, 
Ar$^{2+}$ and Ar$^{3+}$).  {\elecd}({\fsii}) was assumed for all ions. 
All the computations were made with {\sc pyneb}, using the atomic data 
listed in Table~\ref{tab:atomic}.

The He$^+$/H$^+$ abundance ratio was determined using {\sc pyneb}. 
Here the effective recombination coefficients for the {\hei} recombination 
lines calculated by \citet{porteretal12,porteretal13}, who considered the 
effects of collisional contribution and the optical depth in the triplet 
transitions, were adopted.  The final adopted He$^{+}$/H$^+$ ratio is a 
weighted average of the ratios derived from several bright {\hei} lines.

We have detected in the spectrum of Sh~2-100 several RLs of 
heavy-element ions excited by pure recombination (see Fig.~\ref{fig:RLs}). 
We derived the C$^{2+}$/H$^+$ abundance ratio using the measured flux of 
the \cii\ $\lambda$4267 line, \elect({\foiii}), and the Case~B \cii\ 
effective recombination coefficients calculated by \citet{daveyetal00}. 
We have determined the O$^{+2}$/H$^+$ ratio from the intensity of the 
{\oii} Multiplet 1 pure RLs. 
Here the O~{\sc ii} effective recombination coefficients were adopted 
from the Case~B, LS-coupling calculations of \citet{storey94}, and the 
\elect({\foiii}) was assumed.  Under typical physical conditions of 
photoionized nebulae ({\elecd} $\sim$10$^{2}$--10$^{4}$), the relative 
intensities of individual fine-structure lines within the \oii\ M1 
multiplet are not constant but varies as a function of the electron 
density \citep[e.g.,][]{fangliu13}.  This is due to the fact 
that the relative populations of the ground fine-structure levels of the 
recombining ion (e.g., O$^{2+}$ in the case of O~{\sc ii}) deviate from 
the local thermodynamical equilibrium (LTE), which was seldom considered 
in previous atomic calculations.  We considered this deviation from the 
LTE by adopting the prescriptions of \citet{apeimbertpeimbert05} to apply 
appropriate correction of the relative strengths of the \oii\ Multiplet 1 
lines, at densities {\elecd} $<$10$^4$~{\cmc}. 
The derived ionic abundances for our targets, as well the uncertainties, 
are presented tables~\ref{cond_abun_1} and \ref{cond_abun_2}.

As we can see in Table~\ref{cond_abun_1}, the O$^{2+}$/H$^+$ abundance 
ratio derived from the RLs for Sh~2-100 is higher than the abundances 
determined from the CELs.  This is a common observational fact in all 
{\hii} regions where the O$^{2+}$/H$^+$ or C$^{2+}$/H$^+$ ratios can be 
determined using both CELs and RLs.  This is called the abundance 
discrepancy problem and its origin is still under debate 
\citep[e.g.][]{garciarojasesteban07, estebanetal17}. 
For Sh~2-100, the ratio of the two O$^{2+}$/H$^+$ abundances determined 
from RLs and CELs, defined as the abundance discrepancy factor (ADF), is 
0.07~dex, which is one of the lowest values ever found in {\hii} regions. 
This is also slightly lower than the ADF found in the Orion Nebula 
\citep[see][]{estebanetal17}.  Assuming that the abundance discrepancy 
is due to the presence of fluctuations in the spatial distribution of 
{\elect} inside the nebula \citep{torrespeimbertetal80}, we can estimate 
the temperature fluctuation parameter {\tf}, which was first defined by 
\citet{peimbert67} to reconcile the O$^{2+}$/H$^+$ abundance ratios 
determined from RLs and CELs.  The {\tf} value we obtained for Sh~2-100 
is 0.010$\pm$0.010.  We considered two sets of abundances for this object, 
one for the case of {\tf} = 0 and the other for {\tf} = 0.010$\pm$0.010. 
We could not derive {\tf} for the other {\hii} regions in our 
sample.  Therefore their abundances were only calculated for {\tf} = 0.

Since the aim of this paper is to explore the slope of the Galactic 
radial gradient of oxygen in the anticentre direction, we limit our 
study to determine the total abundances of oxygen. In {\hii} regions, 
O is the only element for which no ionization correction factor 
(hereafter ICF) is needed to derive its total abundance; therefore 
its calculation is more accurate than for other elements.  The total 
O abundance is simply the sum of the O$^+$/H$^+$ and 
O$^{2+}$/H$^+$ ratios.  We could also estimate the total abundances 
of He, N, Ne, S, Cl, Ar and Fe for our objects, but we need ICFs, 
which are empirically derived based on the similarity of ionization 
potentials of different ionic species, or estimated from 
photoionization models.  We plan to study the gradients of these 
elements in a future paper where we will carry out a critical 
analysis of the best ICFs for each element, including additional 
observations for the nebulae with accurate determinations of the N 
abundance.  As it has been said before, we detect the very faint 
\cii\ $\lambda$4267 RL in the spectrum of Sh~2-100, and this is 
a remarkable result considering 
the paucity of determinations of C abundance in Galactic {\hii} 
regions. As for the rest of the elements -- apart from O --  we will present 
the C abundance of this object as well as a reassessment of the Galactic gradient of C/H 
in a future paper.  
In Table~\ref{cond_abun_3} we present the recalculated values of 
the O abundances for IC~5146, Sh~2-132, NGC~7635 and Sh~2-156. The 
O/H ratios of M20, M16, M17, M8, NGC~3576, M42, NGC~3603, Sh~2-311 
and NGC~2579 are presented in Table~7 of \cite{estebanetal15}.

\section{The O gradient at the anticentre} 
\label{sec:Ogradient}

As indicated in Section~\ref{sec:intro}, \cite{fernandezmartinetal16} 
recently studied the chemical composition of a sample of {\hii} regions 
in the Galactic anticentre.  We developed the observational program
prior to the publication of that paper.  \cite{fernandezmartinetal16} 
detected auroral lines to derive {\elect} in five objects of their 
sample.  Of those five objects, two are included in our current sample,
Sh~2-83 and Sh~2-212, and a third one, Sh~2-162, corresponds to our 
NGC~7635, for which we recalculated {\elect} and the O/H ratio from 
the emission line ratios measured by \citet{estebanetal16} in their  
deeper spectra taken at the GTC.  The other two objects for which 
\cite{fernandezmartinetal16} determined {\elect} are Sh~2-132 and 
Sh~2-156.  The authors collected spectra of additional {\hii} regions 
taken from the literature and all of them correspond to the observations 
obtained at 2--4\,m telescopes and published between 1979 and 2000. 
When describing the procedure of data collection from the literature, 
\cite{fernandezmartinetal16} stated that they {\it carried out an 
exhaustive bibliographical review} of spectroscopical results of {\hii} 
regions located at $R_{\rm G}$ $>$11~kpc.  It is surprising that they 
overlooked the deepest observations of this kind of objects published 
prior to 2016: the VLT spectroscopy of Sh~2-311 and NGC~2579 published 
by \citet{garciarojasetal05} and \citet{estebanetal13}, respectively. 
Moreover, for Sh~2-311 \cite{fernandezmartinetal16}, instead of using the 
far much better quality data by \citet{garciarojasetal05} adopted the emission 
line ratios obtained by \citet{shaveretal83} with the 3.6\,m telescope 
and the 3.9\,m AAT at La Silla and Siding Spring, respectively.  We 
consider that the new data presented in this paper in combination with 
(a) the compilation of deep spectroscopy from the literature (all of 
them except one obtained from 8-10\,m telescopes) and (b) the 
recalculations of the abundances in a homogeneous way, is a substantial 
improvement in the exploration of the true shape of the O abundance 
gradient at the Galactic anticentre.

\begin{figure*}
 \centering
  \includegraphics[scale=0.82]{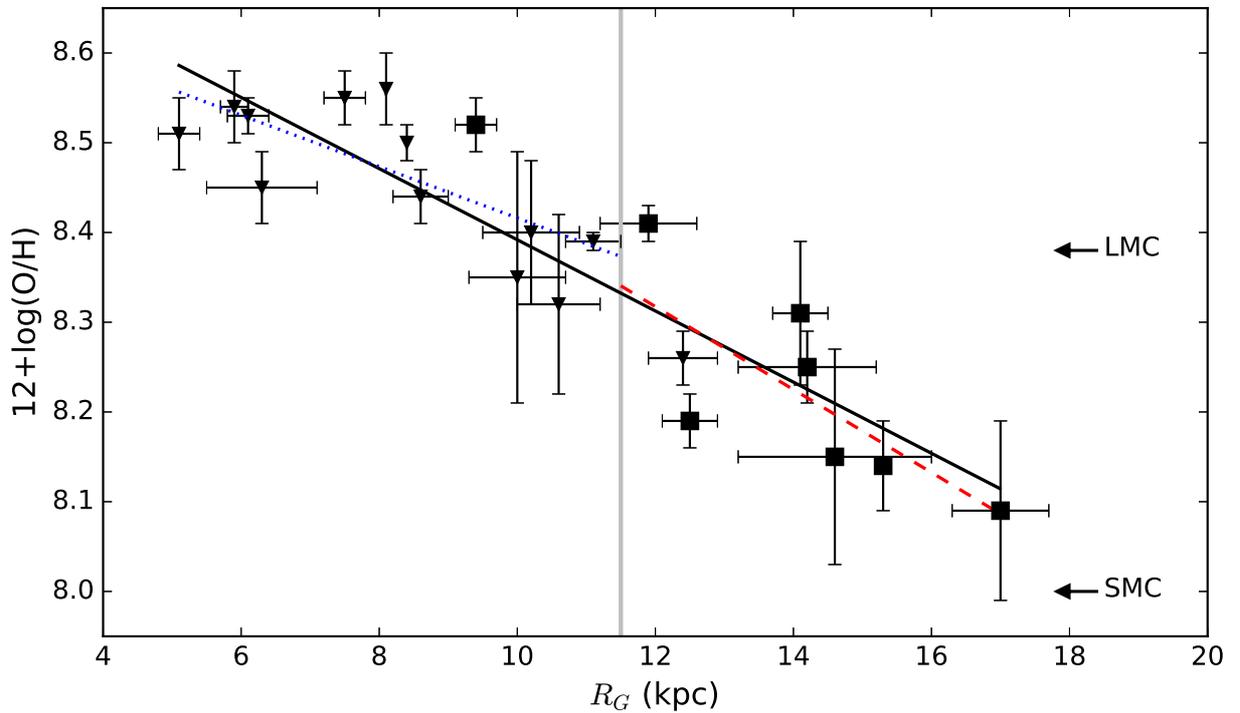}
 \caption{Radial distribution of the oxygen abundance, in logarithm 12+log(O/H), 
 as a function of the Galactocentric distance, $R_{\rm G}$, for the whole sample 
 of Galactic {\hii} regions.  Large filled squares represent the objects of our 
 current sample and small filled triangles are the additional sample.  The solid 
 black line is a least-squares linear fit to all objects.  The vertical 
 grey line marks the position of the isophotal radius of the Milky Way, 
 $R_{25}$ = 11.5~kpc \citep{devaucouleurspence78}.  The blue dotted line is a 
 least-squares linear fit to the {\hii} regions located at $R_{\rm G}$ 
 $<\,R_{25}$, and the red dashed line is the fit to those at $R_{\rm G}$ $>\,R_{25}$. 
 The figure graphically demonstrates that the slope of the gradient does not change 
 significantly across the Galactic disc and the lack of flattening in the outer 
 zone.  The arrows indicate the average O abundances of {\hii} regions in the 
 LMC and SMC \citep{toribiosanciprianoetal17, dominguezguzmanetal17}.}
 \label{fig:Ograd}
\end{figure*}

We performed a least-squares linear fit to the oxygen abundance as 
a function of $R_{\rm G}$ for the whole sample, including objects in 
the Galactic anticentre and the inner disc, i.e., all the objects 
included in tables~\ref{cond_abun_1}, \ref{cond_abun_2} and 
\ref{cond_abun_3} that cover $R_{\rm G}$ 5.1--17~kpc.  The fits give 
the following radial O abundance gradient:
 
\begin{equation}
12 + \log(\mathrm{O/H}) = 8.79(\pm 0.05) - 0.040(\pm 0.005) R_\mathrm{G} ;
\end{equation}

\noindent where the uncertainties are estimated through Monte Carlo 
simulations.  We generated 10$^4$ random values of $R_{\rm G}$ and 
the O abundance for each observational data point assuming a Gaussian 
distribution with a sigma equal to the measurement uncertainty 
of each quantity.  We performed a least-squares linear fit to each of 
these 10$^4$ random distributions.  The uncertainties associated to 
the slope and intercept correspond to the standard deviation of the 
values of these two quantities obtained from the fits.  The spatial 
distribution of the O abundances and the gradient are shown in 
Fig.~\ref{fig:Ograd}. 
Using a similar methodology, \citet{estebanetal15} determined a 
slope of $-$0.043~dex~kpc$^{-1}$ for the O gradient defined by a 
subset of our sample of {\hii} regions restricted to $R_{\rm G}$ 
$\leq$12.4~kpc.  The most recent determinations of the Galactic 
gradient of oxygen for the whole disc available in the literature 
show consistent slopes: \citet{deharvengetal00} obtained a slope of 
$-$0.040($\pm$0.005)~dex~kpc$^{-1}$; \citet{quirezaetal06} derived 
$-$0.043($\pm$0.007)~dex~kpc$^{-1}$; \citet{rudolphetal06} gave 
$-$0.060($\pm$0.010)~dex~kpc$^{-1}$ and $-$0.042($\pm$0.013) 
dex~kpc$^{-1}$, when they used the optical and far-infrared lines, 
respectively; \citet{balseretal11} presented a slope of 
$-$0.045($\pm$0.005)~dex~kpc$^{-1}$. 
As we can see, these determinations are highly consistent, indicating 
that the slope of the radial abundance gradient of oxygen has been 
well established for the Milky Way.

The average difference between the oxygen abundances of 
the {\hii} regions represented in Fig.~\ref{fig:Ograd} and those 
given by the linear fit at their corresponding galactocentric 
distances is $\pm$0.05~dex, similar order of the average uncertainty 
of the abundance determinations.  The maximum difference we find is 
$\pm$0.10~dex.  This is an upper limit of any local inhomogeneity of 
the O abundance that is consistent with our results.

In order to explore whether a change of the slope of the oxygen 
gradient is present in the outer disc, we carried out additional 
least-square fits separately to the {\hii} regions within and beyond 
$R_{25}$ (= 11.5~kpc).  The radial O gradient we found for the 
objects within $R_{25}$ is: 

\begin{equation}
12 + \log(\mathrm{O/H}) = 8.70(\pm 0.07) - 0.029(\pm 0.009) R_\mathrm{G} ;
\end{equation}

\noindent and that for the external {\hii} regions ($R_{\rm G}$ 
$>\,R_{25}$) is:

\begin{equation}
12 + \log(\mathrm{O/H}) = 8.87(\pm 0.23) - 0.046(\pm 0.017) R_\mathrm{G} .
\end{equation}

All these gradients are presented in Fig.~\ref{fig:Ograd}.  As we 
can see in the figure and ascertain by comparing the parameters of 
the gradients, their slopes are consistent within the uncertainties, 
indicating that the shape of the gradient does not change substantially 
across the Galactic disc.  Therefore, our results confirm the absence 
of flattening in the radial O abundance gradient beyond $R_{25}$, 
at least up to $R_{\rm G}$ $\sim$17~kpc or $\sim$1.5$\times$$R_{25}$. 
As a conclusion, we can say that Inside-Out models of galaxy formation 
are also valid to explain the chemical composition of the outer parts 
of the Milky Way.

In Fig.~\ref{fig:Ograd}, we also present the average O abundances 
of the Large and Small Magellanic Clouds (hereafter LMC and SMC) 
obtained by \citet{toribiosanciprianoetal17,dominguezguzmanetal17} from 
deep echelle spectra of a sample of {\hii} regions obtained with the VLT. 
We can see that the ionised gas-phase abundances of O in the 
outer regions (beyond $R_{25}$) of the Milky way is between the 
values measured in the Magellanic Clouds.  Abundances of the 
outermost {\hii} regions are closer to the low value observed in 
the SMC. 
The slope of oxygen gradient obtained by \cite{fernandezmartinetal16} 
for the $R_{\rm G}$ from 11 to 18~kpc is between $-$0.053 and 
$-$0.061~dex~kpc$^{-1}$, consistent with our determinations for the 
outer disc within the uncertainties. 

We have investigated O gradients -- in the Milky Way and other nearby galaxies -- in other papers of our group \citep{estebanetal05, estebanetal13, toribiosanciprianoetal16, toribiosanciprianoetal17}, always focusing the attention 
on the gradient derived from O abundances determined from RLs. One question that we want to briefly address is the possible effect of the abundance discrepancy or the presence of temperature fluctuations in the abundance gradient derived in this 
paper. \citet{garciarojasesteban07} and \citet{estebanetal17} found that {\hii} regions in the discs of spiral galaxies where the ADF has been calculated show quite similar values of such quantity. In fact, \citet{garciarojasesteban07} found that the 
ADF seemed to be independent on some basic properties of {\hii} regions, i. e. metallicity or {\elect}. Moreover, the O/H gradients determined from CELs and RLs are almost identical. Some indication of a possible correlation between the ADF and the O/
H ratio determined from CELs has appeared  with the latest results on low-metalicity objects obtained by \citet{toribiosanciprianoetal17}. This correlation indicates that objects with lower O/H seem to show higher values of the ADF, but this is only apparent for 
objects with 12+log(O/H) $\leq$ 8.1. Taking into account that our objects show O/H ratios -- determined from CELs -- larger than that value, we do not consider this effect may be affecting to the {\hii} regions studied in this paper.

\section{Conclusions}

We present very deep optical spectra of eight {\hii} regions located 
in the anticentre of the Milky Way, with $R_{\rm G}$ between 9.4 and 
17~kpc.  The data were obtained at the 10.4m GTC and the 8.2m VLT. 
We derived \elect({\fnii}) for all the objects and \elect({\foiii}) for six 
of them, this permits to use the direct-{\elect} method based on the measurements 
of the temperature-sensitive auroral lines to derive chemical abundances.  We also included an additional 
sample of 13 {\hii} regions located in the inner and outer disc of the 
Milky Way, whose spectra were also also obtained with large 
telescopes.  Reliable electron temperatures were also determined for 
these additional objects.  The physical conditions and ionic abundances of 
all objects were derived using the same methodology and atomic dataset.  We 
also detected the {\cii} and {\oii} optical recombination lines in 
Sh~2-100, for which we calculated the abundance discrepancy factor in 
O$^{2+}$.  This factor is {higher than unity} but rather small. 

We derived the oxygen abundances for all objects, thus allowing 
determination of the radial abundance gradient over a wide range of $R_{\rm G}$, 
from 5.1 to 17 kpc.  Eight objects are located outside of $R_{25}$, and 13 
inside $R_{25}$.  A least-squares linear fit to the oxygen abundance 
gradient of the whole sample, including the {\hii} regions in the outer and 
inner Galactic disc, gives a slope of $-$0.040$\pm$0.005 dex~kpc$^{-1}$. 
Additional least-squares fits of the inner and outer disk objects, 
separated by $R_{25}$ = 11.5~kpc, give similar slopes.  In particular, the 
slope of the {\hii} regions located beyond $R_{25}$ is $-$0.046$\pm$0.017 
dex~kpc$^{-1}$.  This result indicates that there is no evidence of flattening 
in the radial O abundance gradient beyond $R_{25}$, at least up to 17~kpc 
($\sim$1.5$\times$$R_{25}$).  In general, we find that the scatter in the 
O/H ratios of {\hii} regions across the Galactic disc is not substantially 
larger than the observational uncertainties, with the largest possible 
inhomogeneities of the order of 0.1~dex. 

In an appendix, we explore the radial distribution of \elect({\foiii}) and 
\elect({\fnii}) across the Galactic disk, and found gradients with similar 
positive slopes for both temperatures, with much larger scatter in 
\elect({\fnii}). The shape of the {\elect} gradients is consistent with the absence of flattening in 
the metallicity gradient in the outer Galactic disc.

The results of this 
work indicate that the Inside-Out models of galaxy formation are also 
valid to explain the chemical composition of the outer regions of the 
Milky Way.

\section*{Acknowledgements}

This paper is based on observations made with the Gran Telescopio Canarias (GTC), installed in the Spanish Observatorio del Roque de los Muchachos of the Instituto de Astrof\'isica de Canarias, in the island of La Palma, Spain. We also include previously unpublished data obtained at the European Southern Observatory, Chile, with proposal 070.C- 0008(A). We thank the referee, Grazyna Stasi\'nska for her always wise and useful comments. This work has been funded by the Spanish Ministerio de Econom\'ia y Competividad (MINECO) under project AYA2015-65205-P. JGR acknowledges support from an Advanced Fellowship from the Severo Ochoa excellence program (SEV-2015-0548). LTSC is supported by the FPI Program of the MINECO under grant AYA2011-22614.







\appendix

\section{The {\elect} gradient} 
\label{sec:Tgradient}

The presence of a radial gradient of {\elect} in the Milky Way is well 
established from the radio continuum emission observations 
\citep[e.g.,][]{churchwellwalmsley75,shaveretal83,quirezaetal06} and 
the optical {\foiii} emission line ratios \citep[e.g.,][]{peimbertetal78, 
deharvengetal00}.  Heavy elements such as O or N are the main coolants 
of photoionised gas and therefore the {\elect} of {\hii} regions is 
related to their metallicities.  Consequently, we expect to find 
{\elect} gradients in the Milky Way and in other galaxies. In this paper 
we have derived the \elect({\foiii}) and \elect({\fnii}) for a 
number of {\hii} regions in the outer disc of the Galaxy, and 
compiled/recalculated these temperatures for additional objects 
in the inner Galactic disc.  In total, we obtained 17 determinations of 
the \elect({\foiii}) and 20 \elect({\fnii}), which are presented in 
Tables~\ref{cond_abun_1}, \ref{cond_abun_2} and \ref{cond_abun_3} 
in this paper, and Table~3 in \citet{estebanetal15}.

\begin{figure}
 \centering  
  \includegraphics[width=0.48\textwidth]{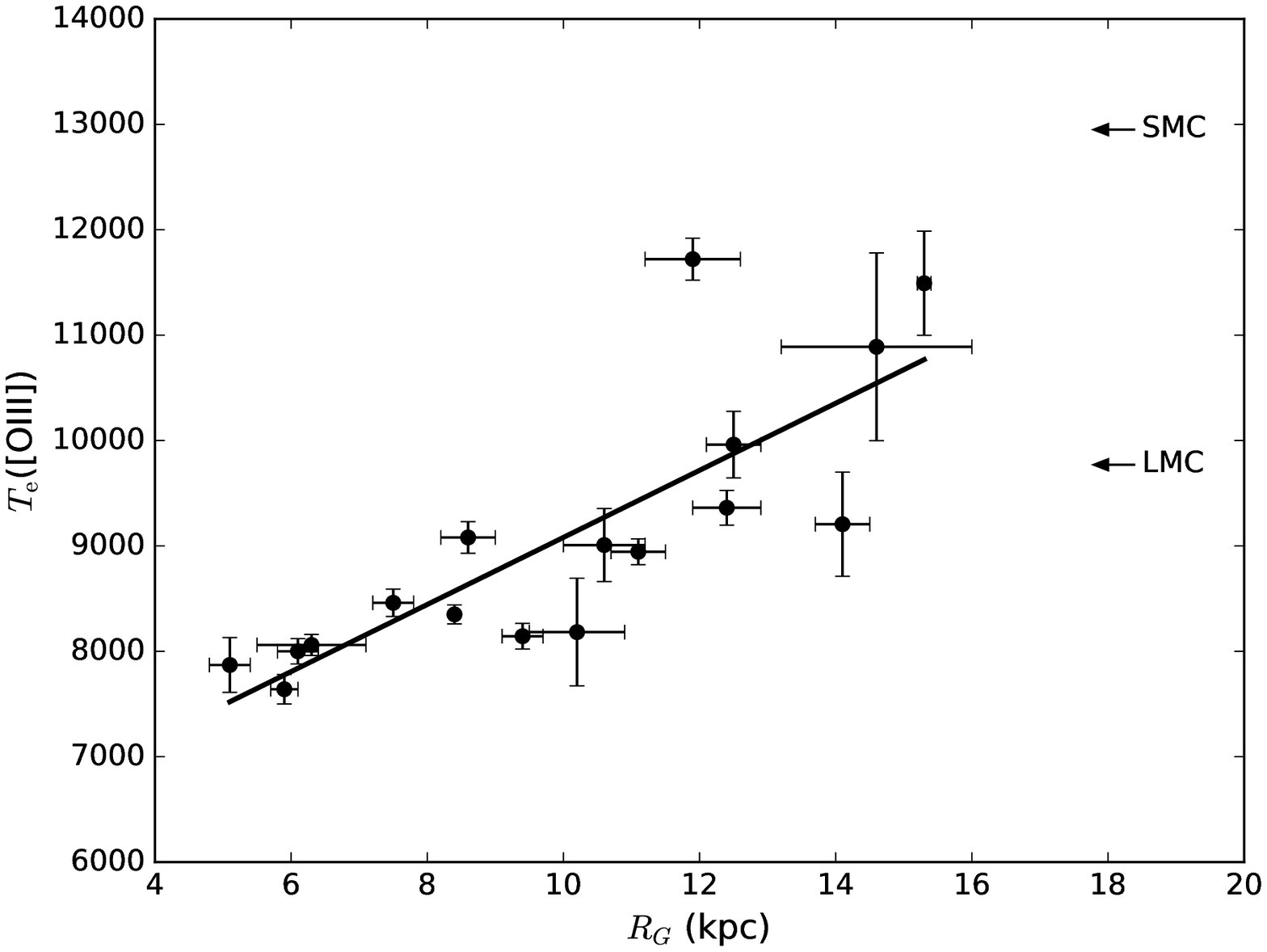}
  \includegraphics[width=0.48\textwidth]{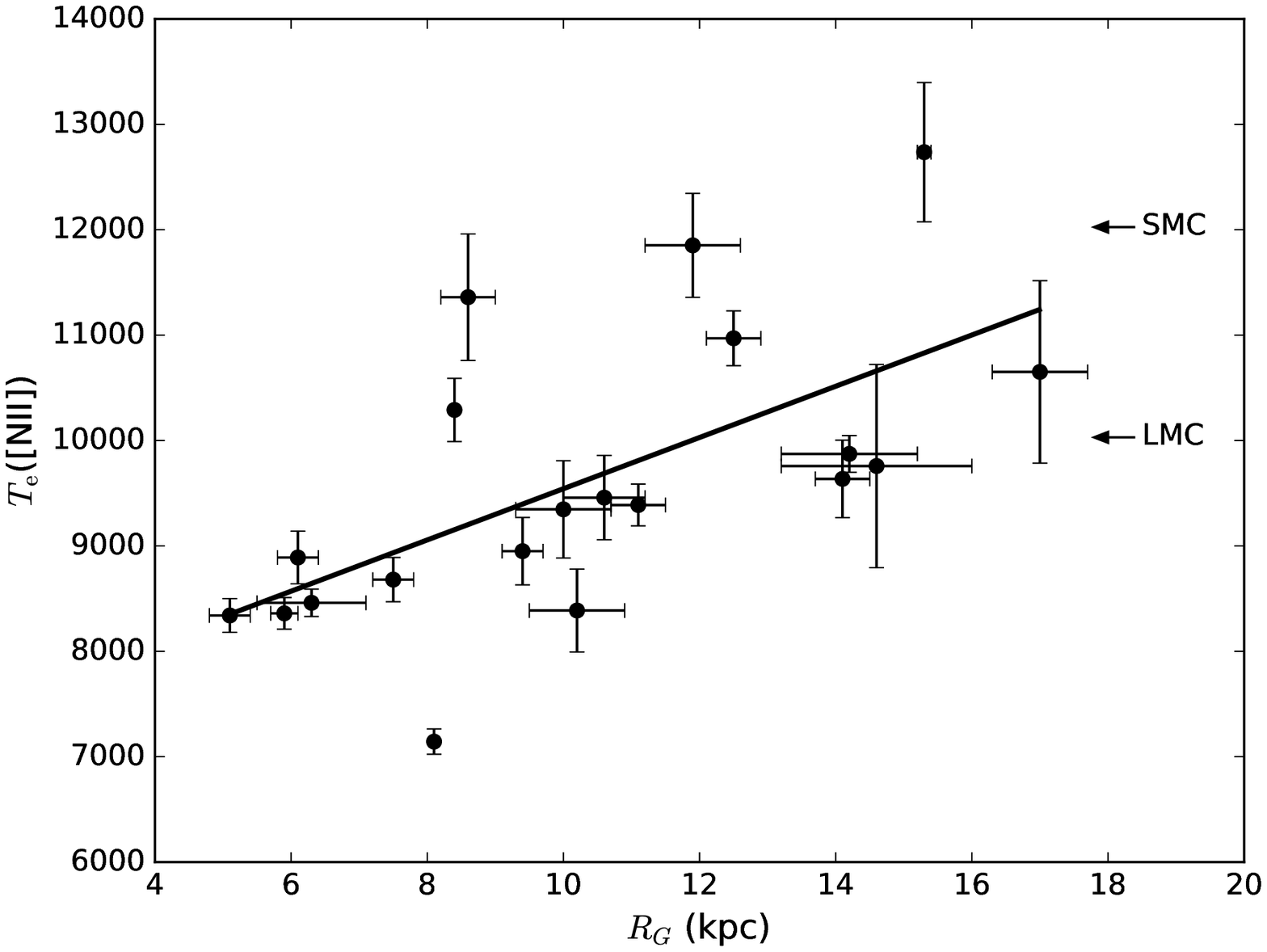}
  \caption{Radial distribution of the \elect({\foiii}) (upper panel) and \elect({\fnii}) 
  (lower panel) for the whole sample of the Galactic {\hii} regions (see the text 
  for description of samples).  The solid black line represents a least-squares linear 
  fit to all objects.  The arrows indicate the average \elect({\foiii}) and 
  \elect({\fnii}) of the {\hii} regions in the LMC and SMC \citep{toribiosanciprianoetal17, 
  dominguezguzmanetal17}.}
 \label{fig:Tgrad}
\end{figure}

In Fig.~\ref{fig:Tgrad} we show the radial distribution of \elect({\foiii}) 
and \elect({\fnii}) of the {\hii} regions of our complete sample.  Our 
least-squares linear fit to the \elect({\foiii}) of the whole sample gives:
\begin{equation}
T_{\mathrm{e}}([\mathrm{O\,III}]) = 5900(\pm 320)\mathrm{K} + 320(\pm 40) \times R_\mathrm{G} ;
\end{equation}

\noindent and our fit to the \elect({\fnii}) gives: 
\begin{equation}
T_{\mathrm{e}}([\mathrm{N\,II}]) = 7110(\pm 360)\mathrm{K} + 240(\pm 40) \times R_\mathrm{G}.
\end{equation}

The derived slope of the \elect({\foiii}) radial gradient, 320$\pm$40 
K~kpc$^{-1}$, is fairly close to the values of 372$\pm$38 K~kpc$^{-1}$ and 287$\pm$46 K~kpc$^{-1}$ determined 
by \citet{deharvengetal00} and \citet{quirezaetal06}, respectively. 
The average difference between the \elect({\foiii}) of the {\hii} regions 
in Fig.~\ref{fig:Tgrad} and the value derived from the linear fit at their 
corresponding distance is 530~K, not much larger than the average uncertainty 
(300~K) of the \elect({\foiii} determinations.  The maximum difference is 
2040~K found in Sh~2-298, an {\hii} region with the hottest ionising star in 
our sample. The ionising source of Sh~2-298 is the Wolf-Rayet (WR) star HD~56925, 
which has been classified as WN4, with $T_{\rm eff}$ $\sim$112,200~K 
\citep{hamannetal06}.  Normal {\hii} regions are usually ionised by OB stars with 
$T_{\rm eff}$ between 30,000 and 45,000~K. 
We have recalculated the average difference between the observed and the fitted 
\elect({\foiii}) of the {\hii} regions by removing the outlier (Sh~2-298) 
and we obtained 430~K, a value closer to a mean observational uncertainty 
of 300~K.  

Fig.~\ref{fig:Tgrad} shows, for the first time, that the \elect({\fnii}) 
of {\hii} regions in the Milky Way also varies with the galactocentric 
distance.  The slope in the \elect({\fnii}), 240$\pm$40 K~kpc$^{-1}$, is 
somewhat lower than that obtained for the \elect({\foiii}) although the 
scatter is considerably larger.
The mean difference between the observed and the fitted \elect({\fnii}) 
of the {\hii} regions presented in Fig.~\ref{fig:Tgrad} is 810~K, higher 
than the average uncertainty (390~K) of all \elect({\fnii}) determinations. 
The exact origin of this scatter is difficult to explain.  N$^+$ is usually
located in the outer regions of the nebulae, and depending on the 
relative development of the Strömgren spheres of the high and low ionization 
species, the differences between \elect({\fnii}) and \elect({\foiii}) may 
vary among different objects.  The maximum difference we 
find in the sample is 2160~K (Sh~2-298) and additional three objects show differences 
of about 2000~K: IC~5146, NGC~3603 and Sh~2-83.  It is interesting 
to note that three out of these nebulae are ionised by stars with effective 
temperatures higher or lower than the usual $T_{\rm eff}$ of O-type 
stars.  IC~5146 is ionised by a B0.5~V star and shows  
the lowest ionization degree among the whole sample. NGC~3603 is the only 
optically visible, giant {\hii} region in the Milky Way and contains several 
WR and hot O3-4 stars \citep{drissen99}. 
As aforementioned in this section, Sh~2-298 is ionised by a WR star 
and also shows a large difference in \elect({\foiii}).  Excluding the four 
objects showing very large differences ($\geq$2000~K) between the 
observed and the fitted \elect({\fnii}), the average difference in 
the rest of the {\hii} regions goes down to 520~K, much closer to the 
typical uncertainty of 390~K. 

\begin{figure}
 \centering  
  \includegraphics[width=0.48\textwidth]{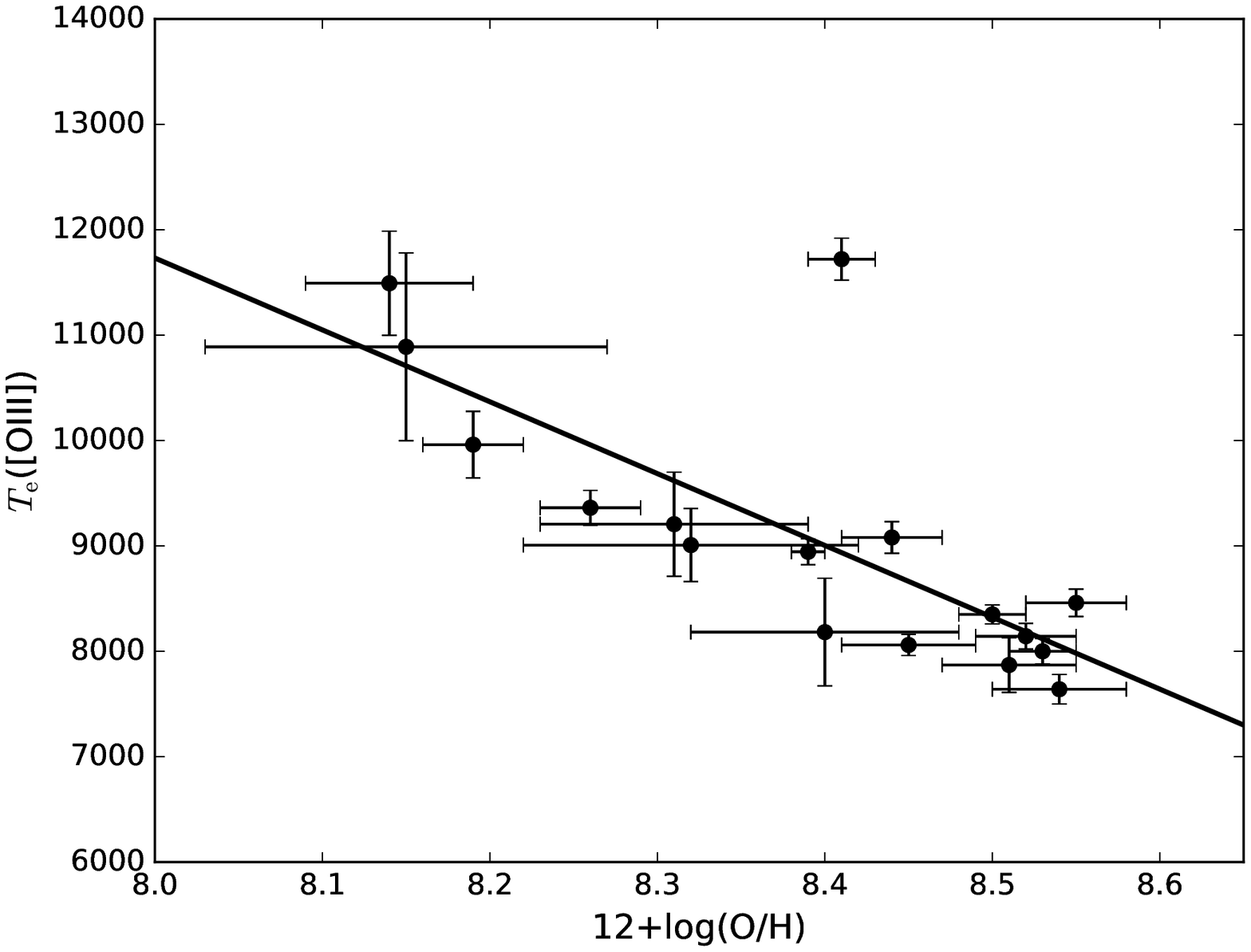}
  \includegraphics[width=0.48\textwidth]{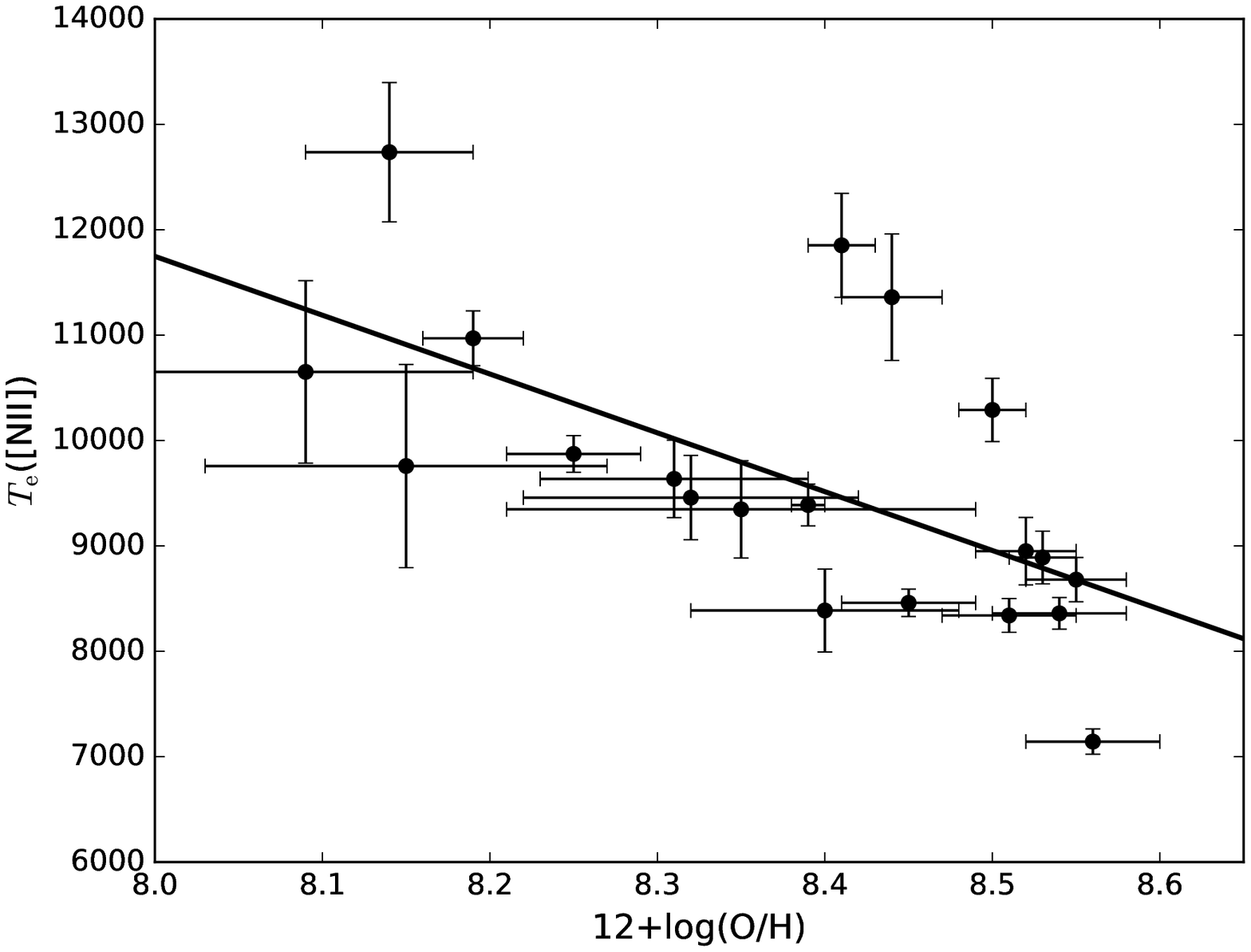}
  \caption{\elect({\foiii}) (upper panel) and \elect({\fnii}) 
  (lower panel) {\it versus} O/H ratio for the whole sample of the Galactic {\hii} regions.  The solid black line represents a least-squares linear 
  fit to all objects. The outlier in the upper panel corresponds to Sh~2-298, ionised by a WR star.}
 \label{fig:TvsO}
\end{figure}

As we can see in Fig.~\ref{fig:Tgrad}, the expected values of 
\elect({\foiii}) and \elect({\fnii}) in {\hii} regions in the outer 
disc of the Milky Way ($R_{\rm G}$ $\sim$17~kpc) are, as in the case of 
the O/H ratios (see Fig.~\ref{fig:Ograd}), between the mean values found 
for the {\hii} regions in the LMC and SMC \citep{dominguezguzmanetal17,
toribiosanciprianoetal17}.  Remarkably, Fig.~\ref{fig:Tgrad} does not give 
any hint about a flattening of the {\elect} gradient in the outer Galactic 
disc, reinforcing the conclusion drawn from Fig.~\ref{fig:Ograd}. 

In Fig.~\ref{fig:TvsO}, we show the dependence of \elect({\foiii}) and \elect({\fnii}) with respect to the O/H ratio. In particular, the correlation 
between \elect({\foiii}) and the O abundance is specially tight. As expected, the objects showing the largest differences from respect the 
linear fit are the same as in Fig.~\ref{fig:Tgrad}. 

The least-squares linear fit to the \elect({\foiii}) {\it versus} O/H ratio relation gives:
\begin{equation}
T_{\mathrm{e}}([\mathrm{O\,III}]) = -15540(\pm 3800)\mathrm{K} - 6820(\pm 1070)\times \mathrm{log(O/H)} ;
\end{equation}
\noindent and to the \elect({\fnii}) {\it versus} O/H ratio relation gives: 
\begin{equation}
T_{\mathrm{e}}([\mathrm{N\,II}]) =- 10580(\pm 4150)\mathrm{K} - 5580(\pm 1160)\times \mathrm{log(O/H)}.
\end{equation}

\begin{figure}
 \centering  
  \includegraphics[width=0.48\textwidth]{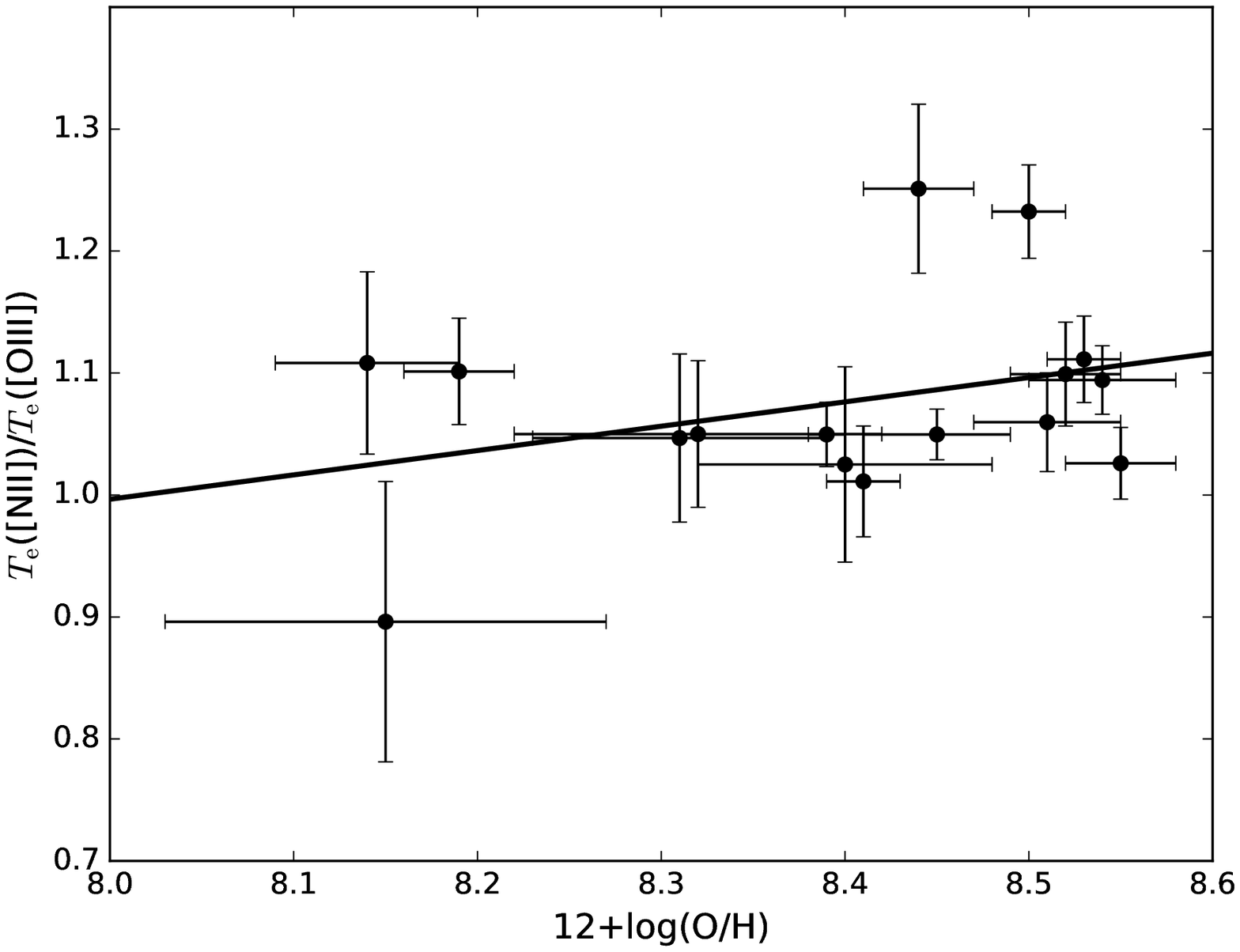}
  \caption{ \elect({\fnii})/\elect({\foiii}) ratio {\it versus} O/H ratio for the whole sample of the Galactic {\hii} regions. The solid black line represents a least-squares linear  fit to all objects.}
 \label{fig:TratiovsO}
\end{figure}

In Fig.~\ref{fig:TratiovsO}, we show the dependence of the \elect({\fnii})/\elect({\foiii}) ratio with respect to the O abundance. The least-squares linear fit of the data included in that figure gives: 
\begin{equation}
T_{\mathrm{e}}([\mathrm{N\,II}])/T_{\mathrm{e}}([\mathrm{O\,III}]) = 1.80(\pm 0.48) +0.20(\pm 0.13)\times \mathrm{log(O/H)} ;
\end{equation}
\noindent that indicates a very small or almost absent correlation between both quantities. 

\begin{figure}
 \centering  
  \includegraphics[width=0.48\textwidth]{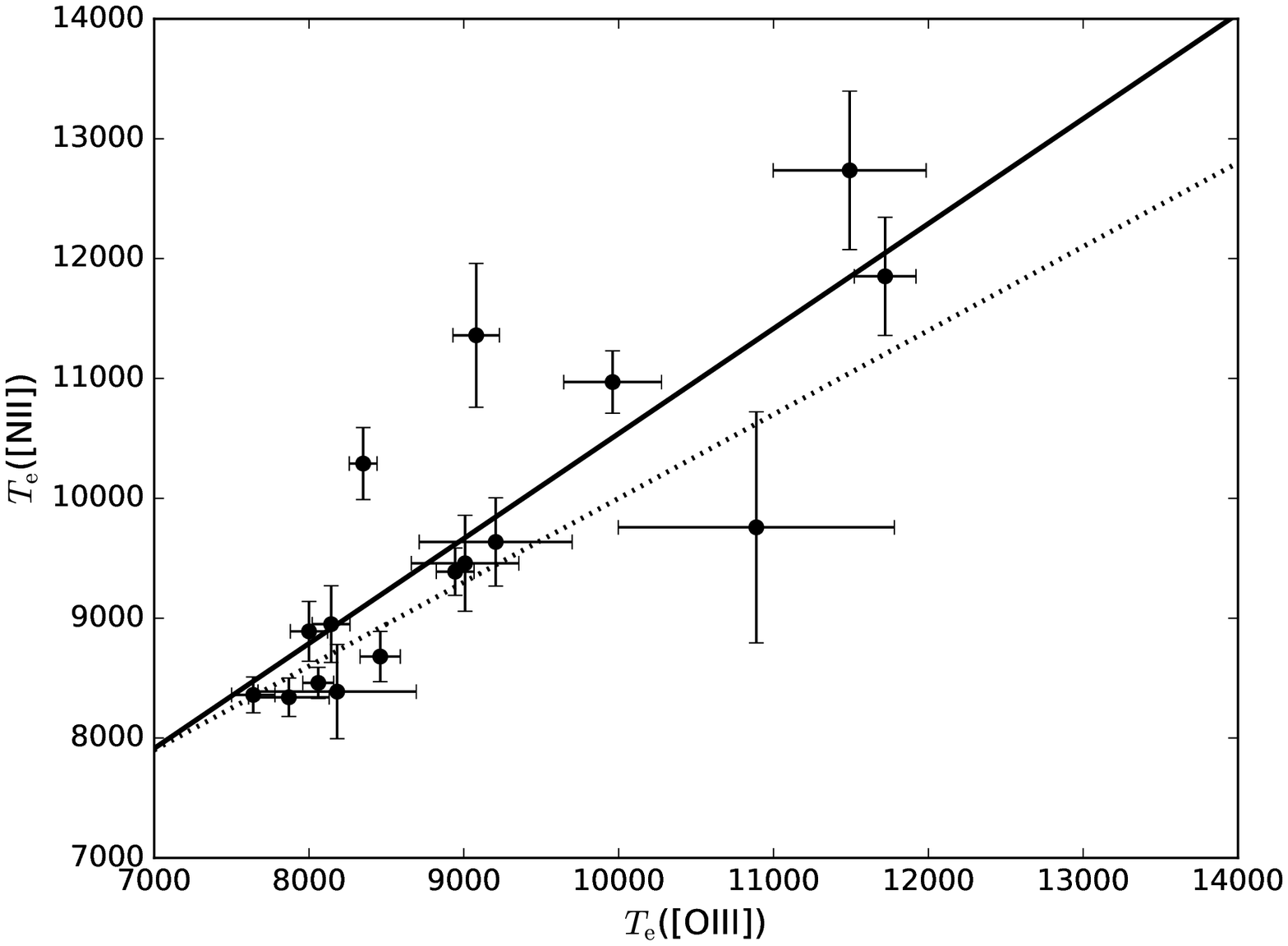}
  \caption{ \elect({\fnii}) {\it versus} \elect({\foiii})  relation for the whole sample of the Galactic {\hii} regions where both quantities have been determined. The solid black line represents a least-squares linear fit to all objects. The dotted line shows the classical relation from \citet{garnett92} given in equation A7.}
 \label{fig:TN_TO}
\end{figure}

Finally, we have obtained the relation between \elect({\fnii}) and \elect({\foiii}) in Fig.~\ref{fig:TN_TO}, finding a relatively tight correlation and the following least-squares linear fit: 
\begin{equation}
T_{\mathrm{e}}([\mathrm{N\,II}]) = 1780(\pm 1280)\mathrm{K} +0.88(\pm 0.15)\times T_{\mathrm{e}}([\mathrm{O\,III}]) ;
\end{equation}
\noindent the slope of the fit is consistent with a 1:1 relation considering the uncertainties. This fit is also very similar to with the classical one obtained by \citet{garnett92} using photoionization models and assuming \elect({\fnii}) = \elect({\foii}): 
\begin{equation}
T_{\mathrm{e}}([\mathrm{N\,II}]) = T_{\mathrm{e}}([\mathrm{O\,II}]) = 3000\mathrm{K} +0.7 \times T_{\mathrm{e}}([\mathrm{O\,III}]). 
\end{equation}
\noindent Our \elect({\fnii})-\elect({\foiii}) relation is also consistent with that obtained by \citet{estebanetal09} from deep spectra for a sample of Galactic and extragalactic {\hii} regions and \citet{valeasarietal16} from the results of an extensive grid of modern photoionization models (see their figure A2).


\bsp	
\label{lastpage}
\end{document}